\documentclass[12pt]{article}
\usepackage[dvips]{graphicx}
\usepackage{epsfig}

\usepackage{latexsym,amsmath,amsfonts,amssymb}
\usepackage[latin1]{inputenc}
\usepackage[american]{babel}
\usepackage{bbm}
\usepackage{comment}

\def\np{Nucl. Phys.}
\def\pl{Phys. Lett.}
\def\prl{Phys. Rev. Lett.}
\def\pr{Phys. Rev.}

\def\im{Invent. Math.}

\def\jhep{J. High Energy Phys.}

\newcommand{\be}{\begin{equation}}
\newcommand{\ee}{\end{equation}}
\newcommand{\beq}{\begin{equation}}
\newcommand{\eeq}{\end{equation}}
\newcommand{\bea}{\begin{eqnarray}}
\newcommand{\eea}{\end{eqnarray}}

\newcommand{\ba}{\begin{eqnarray}}
\newcommand{\ea}{\end{eqnarray}}
\begin{document}
\baselineskip=15.5pt
\pagestyle{plain}
\setcounter{page}{1}
%--------+---------+---------+---------+---------+---------+---------+
%Body

% Ofer's definitions

\def\del{{\partial}}
\def\vev#1{\left\langle #1 \right\rangle}
\def\cn{{\cal N}}
\def\co{{\cal O}}
%\newfont{\Bbb}{msbm10 scaled 1200}     %instead of eusb10
%\newcommand{\mathbb}[1]{\mbox{\Bbb #1}}
\def\IC{{\mathbb C}}
\def\IR{{\mathbb R}}
\def\IZ{{\mathbb Z}}
\def\RP{{\bf RP}}
\def\CP{{\bf CP}}
\def\Poincare{{Poincar\'e }}
\def\tr{{\rm tr}}
\def\tp{{\tilde \Phi}}

\def\TL{\hfil$\displaystyle{##}$}
\def\TR{$\displaystyle{{}##}$\hfil}
\def\TC{\hfil$\displaystyle{##}$\hfil}
\def\TT{\hbox{##}}
\def\HLINE{\noalign{\vskip1\jot}\hline\noalign{\vskip1\jot}}
\def\seqalign#1#2{\vcenter{\openup1\jot
   \halign{\strut #1\cr #2 \cr}}}
\def\lbldef#1#2{\expandafter\gdef\csname #1\endcsname {#2}}
\def\eqn#1#2{\lbldef{#1}{(\ref{#1})}%
\begin{equation} #2 \label{#1} \end{equation}}
\def\eqalign#1{\vcenter{\openup1\jot
     \halign{\strut\span\TL & \span\TR\cr #1 \cr
    }}}
\def\eno#1{(\ref{#1})}
\def\href#1#2{#2}
\def\half{{1 \over 2}}

%--------+---------+---------+---------+---------+---------+---------+
%Hirosi's macros:
\def\ads{{\it AdS}}
\def\adsp{{\it AdS}$_{p+2}$}
\def\cft{{\it CFT}}

\newcommand{\ber}{\begin{eqnarray}}
\newcommand{\eer}{\end{eqnarray}}

\newcommand{\beqar}{\begin{eqnarray}}
\newcommand{\cN}{{\cal N}}
\newcommand{\cO}{{\cal O}}
\newcommand{\cA}{{\cal A}}
\newcommand{\cT}{{\cal T}}
\newcommand{\cF}{{\cal F}}
\newcommand{\cC}{{\cal C}}
\newcommand{\cR}{{\cal R}}
\newcommand{\cW}{{\cal W}}
\newcommand{\eeqar}{\end{eqnarray}}
\newcommand{\tht}{\thteta}
\newcommand{\lm}{\lambda}\newcommand{\Lm}{\Lambda}
\newcommand{\eps}{\epsilon}

%--------+---------+---------+---------+---------+---------+---------+

\newcommand{\nonu}{\nonumber}
\newcommand{\oh}{\displaystyle{\frac{1}{2}}}
\newcommand{\dsl}
   {\kern.06em\hbox{\raise.15ex\hbox{$/$}\kern-.56em\hbox{$\partial$}}}
\newcommand{\id}{i\!\!\not\!\partial}
\newcommand{\as}{\not\!\! A}
\newcommand{\ps}{\not\! p}
\newcommand{\ks}{\not\! k}
\newcommand{\D}{{\cal{D}}}
\newcommand{\dv}{d^2x}
\newcommand{\Z}{{\cal Z}}
\newcommand{\N}{{\cal N}}
\newcommand{\Dsl}{\not\!\! D}
\newcommand{\Bsl}{\not\!\! B}
\newcommand{\Psl}{\not\!\! P}
\newcommand{\eeqarr}{\end{eqnarray}}
\newcommand{\ZZ}{{\rm \kern 0.275em Z \kern -0.92em Z}\;}
%--------------------------------Alfonso's definitions%%%%%%%%%%%%%

% DEFINITIONS

\def\del{{\delta^{\hbox{\sevenrm B}}}} \def\ex{{\hbox{\rm e}}}
\def\azb{A_{\bar z}} \def\az{A_z} \def\bzb{B_{\bar z}} \def\bz{B_z}
\def\czb{C_{\bar z}} \def\cz{C_z} \def\dzb{D_{\bar z}} \def\dz{D_z}
\def\im{{\hbox{\rm Im}}} \def\mod{{\hbox{\rm mod}}} \def\tr{{\hbox{\rm Tr}}}
\def\ch{{\hbox{\rm ch}}} \def\imp{{\hbox{\sevenrm Im}}}
\def\trp{{\hbox{\sevenrm Tr}}} \def\vol{{\hbox{\rm Vol}}}
\def\rl{\Lambda_{\hbox{\sevenrm R}}} \def\wl{\Lambda_{\hbox{\sevenrm W}}}
\def\fc{{\cal F}_{k+\cox}} \def\vev{vacuum expectation value}
\def\nodiv{\mid{\hbox{\hskip-7.8pt/}}}
\def\ie{{\em i.e.}}
\def\ie{\hbox{\it i.e.}}

\def\CC{{\mathchoice
{\rm C\mkern-8mu\vrule height1.45ex depth-.05ex
width.05em\mkern9mu\kern-.05em}
{\rm C\mkern-8mu\vrule height1.45ex depth-.05ex
width.05em\mkern9mu\kern-.05em}
{\rm C\mkern-8mu\vrule height1ex depth-.07ex
width.035em\mkern9mu\kern-.035em}
{\rm C\mkern-8mu\vrule height.65ex depth-.1ex
width.025em\mkern8mu\kern-.025em}}}

\def\RR{{\rm I\kern-1.6pt {\rm R}}}
\def\NN{{\rm I\!N}}
\def\ZZ{{\rm Z}\kern-3.8pt {\rm Z} \kern2pt}
\def\IB{\relax{\rm I\kern-.18em B}}
\def\ID{\relax{\rm I\kern-.18em D}}
\def\II{\relax{\rm I\kern-.18em I}}
\def\IP{\relax{\rm I\kern-.18em P}}
\newcommand{\CS}{{\scriptstyle {\rm CS}}}
\newcommand{\CSs}{{\scriptscriptstyle {\rm CS}}}
\newcommand{\rc}{\nonumber\\}
\newcommand{\bear}{\begin{eqnarray}}
\newcommand{\eear}{\end{eqnarray}}
\newcommand{\W}{{\cal W}}
\newcommand{\F}{{\cal F}}
\newcommand{\x}{{\cal O}}
\newcommand{\LL}{{\cal L}}

\def\mani{{\cal M}}
\def\calo{{\cal O}}
\def\calb{{\cal B}}
\def\calw{{\cal W}}
\def\calz{{\cal Z}}
\def\cald{{\cal D}}
\def\calc{{\cal C}}
\def\to{\rightarrow}
\def\ele{{\hbox{\sevenrm L}}}
\def\ere{{\hbox{\sevenrm R}}}
\def\zb{{\bar z}}
\def\wb{{\bar w}}
\def\nodiv{\mid{\hbox{\hskip-7.8pt/}}}
\def\menos{\hbox{\hskip-2.9pt}}
\def\dr{\dot R_}
\def\drr{\dot r_}
\def\ds{\dot s_}
\def\da{\dot A_}
\def\dga{\dot \gamma_}
\def\ga{\gamma_}
\def\dal{\dot\alpha_}
\def\al{\alpha_}
\def\cl{{closed}}
\def\cls{{closing}}
\def\vev{vacuum expectation value}
\def\tr{{\rm Tr}}
\def\to{\rightarrow}
\def\too{\longrightarrow}

%
% FONTS

%\newfont{\headfont}{cmbx10 scaled 1440}
\newfont{\namefont}{cmr10}
%\newfont{\initialfont}{cmr10 scaled 1200}
\newfont{\addfont}{cmti7 scaled 1440}
\newfont{\boldmathfont}{cmbx10}
%\newfont{\figfont}{cmr7 scaled 1200}
\newfont{\headfontb}{cmbx10 scaled 1728}
%%%%%%%%%%%%%%%%%%%%%%%%%%%%%%%%%%%%%%%%%%%%%%%%%%%%%%%%%%%%%%%%%%%%%%%%%%%%
%%%%%%%%%%%%%%%Stefano and Francesco fonts%%%%%%%%%%%%%%%%%%%%%%%%%%%%%%%%
%%%%%%%%%%%%%%%%%%%%%%%%%%%%%%%%%%%%%%%%%%%%%%%%%%%%%%%%%%%%%%%%%%%%%%%%%%%%
\newcommand{\re}{\,\mathbb{R}\mbox{e}\,}
\newcommand{\hyph}[1]{$#1$\nobreakdash-\hspace{0pt}}
\providecommand{\abs}[1]{\lvert#1\rvert}
\newcommand{\Nugual}[1]{$\mathcal{N}= #1 $}
\newcommand{\sub}[2]{#1_\text{#2}}
\newcommand{\partfrac}[2]{\frac{\partial #1}{\partial #2}}
\newcommand{\bsp}[1]{\begin{equation} \begin{split} #1 \end{split} \end{equation}}
\newcommand{\calF}{\mathcal{F}}
\newcommand{\calO}{\mathcal{O}}
\newcommand{\calM}{\mathcal{M}}
\newcommand{\calV}{\mathcal{V}}
\newcommand{\bbZ}{\mathbb{Z}}
\newcommand{\bbC}{\mathbb{C}}

\numberwithin{equation}{section}

\newcommand{\Tr}{\mbox{Tr}}    % trace over gauge indices

%%%%%%%%%%%%%%%%%%%%%%%%%%%%%%%%%%%%%%%%%%%%%%%%%%%%%%%%%%%%%%%%%%%%%%%%%%%%
%%%%%%%%%%%%%%%%%%%%%%%%%%%%%%%%%%%%%%%%%%%%%%%%%%%%%%%%%%%%%%%%%%%%%%%%%%%%

%
\renewcommand{\theequation}{{\rm\thesection.\arabic{equation}}}
\begin{titlepage}
%
%\rightline{US-FT-3/07}
\vspace{0.1in}

\begin{center}
\Large \bf Holographic flavor in ${\cal N}=4$ gauge theories in 3d from wrapped branes
\end{center}
\vskip 0.2truein
\begin{center}
Alfonso V. Ramallo\footnote{alfonso@fpaxp1.usc.es},
Jonathan P.  Shock\footnote{shock@fpaxp1.usc.es} and 
Dimitrios Zoakos\footnote{zoakos@fpaxp1.usc.es}\\
\vspace{0.2in}
\it{
Departamento de  F\'\i sica de Part\'\i culas, Universidade
de Santiago de
Compostela\\and\\Instituto Galego de F\'\i sica de Altas
Enerx\'\i as (IGFAE)\\E-15782, Santiago de Compostela, Spain}

\vspace{0.2in}
\end{center}
\vspace{0.2in}
%\begin{center}
\centerline{{\bf Abstract}}
We study the addition of flavor to the gravity dual of ${\cal N}=4$ three-dimensional gauge theories obtained by wrapping $N_c$ D4-branes on a two-cycle of a non-compact Calabi-Yau two-fold.  In this setup the flavor is introduced by adding another set of  D4-branes that are extended along the non-compact directions of the Calabi-Yau which are normal to the cycle which the color branes wrap. The analysis is performed both in the quenched and unquenched approximations. In this latter case we compute the backreacted metric and we show that it reproduces the running of the gauge coupling. The meson spectrum and the behavior of Wilson loops are also discussed and the holographic realization of the Higgs branch is analyzed. Other aspects of this system studied are the entanglement entropy and the non-relativistic version of our backgrounds.

\smallskip
\end{titlepage}
\setcounter{footnote}{0}

\tableofcontents

%--------+---------+---------+---------+---------+---------+---------+
%Body

\newpage

\section{Introduction}

The gauge/gravity correspondence \cite{Maldacena:1997re} is one of the major achievements of string theory in the last ten years (see  \cite{MAGOO} for a review). Its extension to include more realistic theories is, clearly, a topic of great interest. In particular, the addition of matter degrees of freedom is essential to get nearer to the string theory description of QCD-like theories. This addition can be performed by including extra (flavor) branes \cite{KK}, which should be extended along all the gauge theory directions and  wrap a non-compact cycle in order to make its worldvolume symmetry a global symmetry from the point of view of the gauge theory living in the  color branes. They introduce an open string sector that corresponds to having hypermultiplets transforming in the fundamental representation in the gauge theory side.

 If the number of flavors, $N_f$, is small compared with the number of colors, $N_c$, one can treat the flavor branes as probes in the supergravity background created by the color branes.  This defines the so-called quenched approximation, which corresponds, in the field theory side, to neglecting quark loops, which are suppressed  by factors $1/N_c$ in the 't Hooft large $N_c$ expansion \cite{'tHooft:1973jz}. This quenched holographic flavor has been explored extensively in the past few years.  In particular, by analyzing the normalizable fluctuations of the probe branes, the spectra of mesonic excitations of different theories have been analyzed (see \cite{Erdmenger:2007cm} for a review and a list of references).

On the contrary, if the number of flavors is of the order of the number of colors ($N_f\sim N_c$) the backreaction of the flavor branes on the geometry cannot be ignored and one has to deal with a system of gravity plus branes, the latter acting as dynamical sources for the different supergravity fields.  On the field theory side the inclusion of the backreaction in this $N_f\sim N_c$ regime is equivalent to considering the so-called Veneziano limit \cite{Veneziano:1976wm}, in which $N_c$ and $N_f$ are large and their ratio $N_f/N_c$ is fixed. In this limit quark loops are not suppressed and the flavor is unquenched.

Another direction in which the original gauge/gravity duality has been generalized is by  extending it to theories with lower amounts of  supersymmetry. A general strategy to carry out  this extension of the correspondence  to less supersymmetric  models is to consider higher dimensional branes wrapping cycles. At energies small compared with the size of the cycle the theory becomes effectively four-dimensional. Moreover, the gauge theory living on the worldvolume of the wrapped brane has to be topologically twisted in order to preserve some fraction of supersymmetry.   Examples of duals of $4d$ gauge theories constructed in this way are the geometries analyzed in \cite{CVMN} for ${\cal N}=1$ and in  \cite{N2d4} for ${\cal N}=2$, which correspond to D5-branes wrapping a two-cycle inside a Calabi-Yau (CY) manifold.

In this paper we will study the dual of ${\cal N}=4$ gauge theories in three space-time dimensions, obtained by wrapping D4-branes of the type IIA theory on a two-cycle of a Calabi-Yau two-fold. The corresponding unflavored supergravity solution was found in \cite{Maldacena:2000mw} and studied in detail in \cite{Divecchia}, where it was shown to reproduce the exact perturbative running coupling constant and the metric of the moduli space of the gauge theory. In the present paper we will analyze the addition of flavor branes in this setup. These flavor branes are also D4-branes, which are extended along non-compact directions of the CY two-fold in such a way that no further supersymmetry is broken. 

We will analyze the addition of flavor to the $3d$, ${\cal N}=4$ theory both in the quenched and unquenched approaches. In the quenched formalism we will compute numerically  the spectrum of mesonic excitations and we will be able to find a very good analytical estimate of the masses.  We will  also study the Wilson loops and the corresponding quark-antiquark potentials. In this latter case we will discover that a phase transition is produced when some of the parameters of the solution are varied and we will compute the corresponding critical exponents.

We will include the backreaction by following the proposal of ref. \cite{Casero:2006pt}, in which the localized brane sources are substituted by a continuous distribution of flavor branes in their transverse directions (see also \cite{noncritical} for a similar analysis in the context of non-critical string theory). This smearing brane approach has been applied successfully to various string duals in several dimensions (\cite{Casero:2007pz}-\cite{Gaillard:2008wt}). It has the advantage that it avoids having the $\delta$-function sources of the localized branes in the BPS equations  and, therefore, makes searching for solutions more feasible.

In our case we will start with the same ansatz for the metric as in the unflavored system.
However,  the presence of flavor branes modifies the Bianchi identity of the RR field strength $F_4$, which forces us to modify the unflavored ansatz for $F_4$. Once this fact is taken into account, one can get a system of first-order differential equations by imposing that the supersymmetric variation of the gravitino and dilatino of type IIA supergravity vanish. These first-order equations are simple, but they are difficult to solve in general. However, we will see that they can be solved analytically in certain regions of the space. Remarkably, this is enough to compute the modification of the running of the coupling constant due to the matter hypermultiplets in the gravity solution. We will show that this modification matches exactly the field theory results. 

The organization of the rest of this paper is the following.  We will start in section \ref{unflavored} by reviewing the unflavored background. We  will set up our notations and  write the BPS equations  and their solution for the unflavored system. In section \ref{flavoredbackground} we study the addition of flavor branes. In particular, we obtain the BPS equations for the backreacted background, which are then numerically integrated. Section \ref{gaugetheory} is devoted to analyzing  the matching between our flavored background and its field theory dual. By means of a probe calculation we verify that the running of the gauge coupling with flavor is reproduced by our solution. We also discuss in this section how to realize the Higgs branch in our setup. 

The analysis of the meson spectrum is the object of section \ref{mesons}. We study the mass levels both in the quenched and unquenched solutions and we also consider the meson spectra  in the Higgs branch.  In section \ref{Wilsonloops} we explore the behavior of the Wilson loops. In our study of the energy of the quark-antiquark pair we will find some critical phenomena and we will evaluate the corresponding critical exponents. Section \ref{conclusions} is devoted to presenting our conclusions and  summarizing our results. 

The paper is completed with several appendices. In appendix \ref{BPSeqs} we give details of the derivation of the BPS equations for the general flavored system and we check that the equations of motion of the system are satisfied if the BPS equations hold. In appendix \ref{embeddings} we find the supersymmetric embeddings of D4-brane probes in our background. In appendix \ref{additonal-backgrounds} we find additional solutions of the unflavored equations and we obtain the background dual to  a non-relativistic system that can be generated by deforming our solutions. Finally, in appendix \ref{additonal-backgrounds} we analyze the entanglement entropy for our model in the UV.

\section{The unflavored background}
\label{unflavored}

Following the analysis of refs. \cite{Maldacena:2000mw} and \cite{Divecchia}, let us
consider the background of type IIA supergravity created by a
stack of $N_c$ D4-branes wrapped on a two-cycle ${\cal C}_2$  of a Calabi-Yau cone of complex dimension two,  according to the following brane setup
\begin{center}
\begin{tabular}{|c|c|c|c|c|c|c|c|c|c|c|}
\multicolumn{4}{c}{ }&
\multicolumn{4}{c}{$\overbrace{\phantom{\qquad\qquad\qquad}}^{\text{CY}_2}$}\\
\hline
&\multicolumn{3}{|c|}{$\mathbb{R}^{1,2}$}
&\multicolumn{2}{|c|}{$S^2$}
&\multicolumn{2}{|c|}{$N_2$}
&\multicolumn{3}{|c|}{$\mathbb{R}^{3}$}\\
\hline
D$4$ &$-$&$-$&$-$&$\bigcirc$&$\bigcirc$&$\cdot$&$\cdot$&$\cdot$&$\cdot$&$\cdot$\\
\hline
\end{tabular}
\end{center}
where $S^2$ represents the directions of the two-cycle (which is topologically a two-sphere) and $N_2$ are the directions of the normal bundle to ${\cal C}_2$ .
In the above setup a circle represents wrapped directions, whereas the symbols ``-" and ``."  denote unwrapped worldvolume  and transverse directions respectively. We shall parameterize the cycle ${\cal C}_2$ by means of two angular coordinates $(\tilde\theta, \tilde\phi)$   with $0\le \tilde\theta <\pi$ and $0\le\tilde\phi<2\pi$ and we will denote by $\sigma$ the radial coordinate of the CY cone which, together with an angular coordinate $\psi$, will parameterize the normal bundle $N_2$.  Moreover, we shall choose a system of spherical coordinates for the transverse $\mathbb{R}^{3}$, $r$ being the corresponding radial coordinate and $(\theta, \phi)$ the angular variables ($0\le \theta <\pi$, $0\le\phi<2\pi$). The concrete ansatz for the ten-dimensional string frame metric we will adopt is the following:
\begin{eqnarray}
ds_{st}^2&=&e^{2\Phi}\left[dx^2_{1,2}+{\cal Z} R^2\left(d\tilde{\theta}^2+\sin^2\tilde{\theta}d\tilde{\phi}^2\right)\right]\,+\,\nonumber\\
&+&e^{-2\Phi}\left[\frac{1}{{\cal Z}}\left(d\sigma^2+\sigma^2\left(d\psi+\cos\tilde{\theta}d\tilde{\phi}\right)^2\right)+
dr^2+r^2\left(d\theta^2+\sin^2\theta d\phi^2\right)
\right] \ ,
\label{metric-ansatz}
\end{eqnarray}
where $dx^2_{1,2}$ denotes the Minkowski metric in 2+1 dimensions and the range of $\psi$ is $0\le \psi<2\pi$. Notice that $\psi$ is fibered over the $(\theta, \phi)$ two-sphere. For convenience we have included in (\ref{metric-ansatz}) the radius $R$, given by:
\beq
R^3\,=\,8\pi g_s\,N_c\,(\alpha')^{{3\over 2}}\,\,,
\label{R}
\eeq
with $g_s$ and $\alpha'$ being respectively the string coupling constant and the Regge slope. The ansatz (\ref{metric-ansatz}) contains two functions: ${\cal Z}$, which controls the size of the cycle, and $\Phi$, which is the dilaton of the type IIA theory. Both of them should be considered as functions of the two radial coordinates $r$ and $\sigma$:
\be
\Phi\,=\,\Phi(r,\sigma)\,\,,\qquad\qquad
{\cal Z}\,=\,{\cal Z}(r,\sigma)\,\,.
\ee
As in any other background generated by D4-branes, the ansatz should be endowed with an RR four-form $F_4$. Let $C_3$ denote the corresponding three-form potential ($F_4=dC_3$). We shall adopt the following ansatz for $C_3$:
\be
C_3\,=\, - \, g\,\omega_2 \, \wedge \, (d\psi+\cos\tilde{\theta} \,d\tilde{\phi}) \, ,
\label{C3}
\ee
where $g(r,\sigma)$ is a new function and $\omega_2$ is the volume element
of the $(\theta,\phi)$ two-sphere:
\begin{equation}
\omega_2 \, = \, \sin \theta \, d\theta \, \wedge \, d\phi \, .
\label{omega2}
\end{equation}
The corresponding RR four-form field strength will be:
\begin{equation}\label{F4-form}
F_4\,=\,-\,(\dot{g}\,d\sigma\,+\,g' \, dr)\,\wedge \, \omega_2 \, \wedge \, (d\psi+\cos\tilde{\theta} \,d\tilde{\phi})\,+\,g\,\,\tilde{\omega}_2 \, \wedge \, \omega_2 \, ,
\end{equation}
where we  have denoted:
\be
'\,\equiv\,\partial_{r}\,\,,\qquad\qquad
\dot{}\,\equiv\,\partial_{\sigma} \, .
\ee

We will require that our background preserves eight supersymmetries. which is the appropriate number of SUSYs for a supergravity dual of ${\cal N}=4$ gauge theories in three dimensions. As shown in detail in appendix \ref{BPSeqs}, the vanishing of the different components of the supersymmetric variations of the  gravitino and dilatino gives rise to a system of first-order BPS equations for the functions $\Phi$, ${\cal Z}$ and $g$ entering our ansatz. This system is the following:
\begin{eqnarray}
& g\,= \, - R^{2}\,r^2\,{\cal Z'} \ , \quad \quad
\,e^{-4\Phi}\, \sigma\,=\,R^{2}\,{\cal Z\dot{Z}} \ ,
\nonumber
\\ \nonumber
\\ & g'\,=\,-4\,\sigma\,r^2\,e^{-4\Phi}\dot{\Phi} \ , \quad \quad
\dot g\,=\,\,-\sigma\,R^{-2}\,{\cal
Z}^{-2}\,e^{-4\Phi}\,g\,+\,4\,\sigma\,r^2\,{\cal
Z}^{-1}\,e^{-4\Phi}\,\Phi' \, . &
 \label{no-flavor-BPSsystem} 
\end{eqnarray}
It is interesting to notice that not all the equations in (\ref{no-flavor-BPSsystem}) are independent. Actually, one can check that the equation for $\dot g$ in (\ref{no-flavor-BPSsystem}) can be obtained from the others. Moreover, one can combine the different equations in (\ref{no-flavor-BPSsystem}) and get a single second-order PDE for the function ${\cal Z}(r, \sigma)$, namely:
\be r\,{\cal Z}\,\left(\,\dot{{\cal Z}}\,-\,\sigma\, \ddot{{\cal
Z}}\,\right)\,=\,\sigma\,\left(\,r\,\dot{{\cal Z}}^2\,+\, r\,{\cal
Z}''\,+\, 2 {\cal Z}'\,\right)\,\,. \label{PDE-z-no-flavored} 
\ee
Notice that, if ${\cal Z}$ is known, the other functions $\Phi$ and $g$ can be determined  from the first two equations in (\ref{no-flavor-BPSsystem}). Moreover, we check in appendix \ref{BPSeqs} that the second order equations of motion for the RR four-form $F_4$, dilaton $\Phi$ and the metric $G_{MN}$ of type IIA supergravity follow from the system  (\ref{no-flavor-BPSsystem}).

%
%%%%%%%%%%%%%%%%%%%%%%%%%%%%%%%%%%%%%%%%%%%%%%%%%%%%%%%%%%%%%%%%%%%%%%%%%%

\subsection{Integration of the BPS system}
\label{unflavored-integration}

The BPS system (\ref{no-flavor-BPSsystem}) can be integrated by elementary methods when $\sigma=0$ and $r$ varies. Indeed, it follows from the last line in  (\ref{no-flavor-BPSsystem}) that $ g(r, \sigma=0)$ is constant. Let us put:
\be g(r, \sigma=0)=g_0\,\, . \ee
Then, the first equation in \eqref{no-flavor-BPSsystem}
for $\sigma=0$ can be readily integrated, namely:
\be 
{\cal Z}'(r, 0)\,=\,-\,{g_0\over r^2\,R^2} \quad \Rightarrow
\quad {\cal Z}(r, 0)\,=\,{g_0\over r\,R^2}\,+\,{\rm constant} \, . 
\label{Z-sigma0}
\ee
The actual value of $g_0$ can be obtained from the quantization condition of the RR four-form flux $F_4$. Actually, the best way to perform this analysis is by using the approach in which the solution is obtained by uplifting from gauged supergravity. This study was done in refs. \cite{Maldacena:2000mw, Divecchia} and allows one to find a solution of the system (\ref{no-flavor-BPSsystem}) for arbitrary values of the variables $r$ and $\sigma$. Here we will just reproduce this solution with our notations. First of all, let us define the function $\Gamma({\cal Z})$ as follows:
\beq
\Gamma({\cal Z})\equiv{\cal Z}_{*}\,+\,2({\cal Z}-{\cal Z}_{*})\,+\,{16\kappa\over {\cal Z}_{*}}\,
({\cal Z}-{\cal Z}_{*})^2\,\,,
\label{GammaZ}
\eeq
where ${\cal Z}_{*}$ and $\kappa$ are constants.
Then, ${\cal Z}(r, \sigma)$ is determined implicitly as the solution of the equation:
\beq
({\cal Z}_{*}-{\cal Z})\,\Big[\,r^2\,+\,{\sigma^2\over \Gamma({\cal Z})}\,\Big]^{{1\over 2}}\,=\,{R\over 8}\,\,.
\label{Zimplicit}
\eeq
We are able to solve this quartic equation in ${\cal Z}$ exactly, although we must be careful that when we use this solution in the following we are always picking up the appropriate root. For each value of $r$ and $\sigma$ it must be checked that the solution is real and less than ${\cal Z}_{*}$.
Notice from (\ref{Zimplicit}) that  ${\cal Z}_{*}$ is just the constant value approached by  ${\cal Z}$ in the UV, {\ie} when either $r$ or $\sigma$ is large. 
Moreover, $g(r, \sigma)$ and $\Phi(r, \sigma)$ are given by:
\beq
g\,=\,{R^2\,({\cal Z}-{\cal Z}_{*})\,r^3\over 
r^2\,+\,{{\cal Z}\over \Gamma^2({\cal Z})}\,\sigma^2}\,\,,\qquad\qquad
e^{-4\Phi}\,=\,R^2\,{{\cal Z}\,(\,{\cal Z}_{*}-{\cal Z}\,)\over 
\Big[\,r^2\,+\,{{\cal Z}\over \Gamma^2({\cal Z})}\,\sigma^2\Big]\,\Gamma({\cal Z})}\,\,.
\label{g-Phi}
\eeq
As a check of this solution, one can easily prove by taking derivatives of (\ref{GammaZ})-(\ref{Zimplicit}) that the BPS system (\ref{no-flavor-BPSsystem}) is satisfied. Moreover, by taking $\sigma=0$ in (\ref{Zimplicit}) one can verify that $ {\cal Z}(r, 0)$ is indeed of the form  (\ref{Z-sigma0}), with $g_0$  being given by:
\be 
 g_0\,=- {R^3 \over 8} \ . 
\ee
Using this result we can rewrite  ${\cal Z}(r, 0)$ as:
\be
 {\cal Z}(r, 0)\,=\,{\cal Z}_{*}\,-{R \over 8\,r} \,\,.
\label{z-sigma0}
 \ee
Furthermore, by using (\ref{z-sigma0}) to evaluate the right-hand side of the second equation in (\ref{g-Phi}), one can obtain the value of the dilaton at $\sigma=0$, namely:
\beq
e^{-4\Phi(r,0)}\,=\,{{\cal Z}_*\,-\,{R\over 8r}\over
{\cal Z}_*\,-\,{R\over 4r}\,+\,{\kappa\over 4 {\cal Z}_*}\,{R^2\over r^2}}\,\,\,
{R^3\over r^3}\,\,.
\label{dilaton-at-sigma0}
\eeq
From the explicit expressions for ${\cal Z}$ and $\Phi$ written above one easily concludes that, when $r$ and $\sigma$ are small enough, the supergravity solution is not valid because the function ${\cal Z}$ becomes negative and/or the dilaton $\Phi$ becomes complex. This phenomenon is related to the so-called enhan\c{c}on mechanism \cite{Johnson:1999qt} (see below). We can estimate the scale at which this mechanism occurs by  computing from (\ref{z-sigma0}) the value of $r$ for which ${\cal Z}(r,0)$ vanishes. This determines the so-called enhan\c{c}on radius $r_e$, given by:
\beq
r_e\,=\,{R\over 8 {\cal Z}_*}\,\,.
\label{re}
\eeq
Notice also that the sign of the right-hand side of (\ref{dilaton-at-sigma0}) becomes negative for sufficiently small $r$. Actually, the numerator in (\ref{dilaton-at-sigma0}) changes its sign precisely at $r=r_e$ , whereas the change of sign of the denominator depends on the value of the constant $\kappa$. Indeed, by analyzing the discriminant of the quadratic function in the denominator of (\ref{dilaton-at-sigma0}) one easily concludes that this equation has no real roots for $\kappa>1/16$. Thus in this case 
$e^{-4\Phi(r,0)}$ changes its sign precisely at the enhan\c{c}on radius $r_e$. However, for
$\kappa<{1\over 16}$ this change of sign occurs for larger values of $r$, namely for $r=r_H$, where $r_H$ is given by:
\beq
r_H\,=\,\big(\,1\,+\sqrt{1-16\kappa}\,\big)\,\,r_e\,\,.
\label{rH}
\eeq
Clearly, when $\kappa\le 1/16$ the space ends, at $\sigma=0$, when $r=r_H\ge r_e$.

\subsection{UV form of the metric}
As mentioned above, it follows from (\ref{Zimplicit}) that the function ${\cal Z}$ approaches the constant value ${\cal Z}={\cal Z}_*$ as one moves into the UV region. 
 Since the solution of  the  algebraic equation  (\ref{Zimplicit}) for ${\cal Z}(r, \sigma)$ is complicated, we can try to perform  an expansion around this constant value. Keeping the first non-trivial term, ${\cal Z}(r, \sigma)$ can be approximated by the following expression:
\begin{equation} \label{UV-Z}
{\cal Z}(r, \sigma) \thickapprox {\cal Z}_{*} - \frac{R}{8} \frac{{\cal Z}_{*} ^{1/2}}{\sqrt{r^2{\cal Z}_{*}+\sigma^2}} \ ,
\end{equation}
while for the dilaton $\Phi$ we have:
\begin{equation} \label{UV-H}
e^{-4\Phi(r, \sigma)} \thickapprox \frac{R^3}{8} \frac{ {\cal Z}_{*}^{3/2}}{\left[ \sqrt{r^2{\cal Z}_{*}+\sigma^2} \right]^3}
\ .
\end{equation}
Notice that the expression for ${\cal Z}$ in \eqref{UV-Z} gives the exact result \eqref{z-sigma0} for $\sigma=0$. The above analysis suggests that in the UV region the combination $r^2{\cal Z}_{*}+\sigma^2$ plays a significant role. Having this in mind we define a new set of variables, $u$ and $\hat{\alpha}$, as follows:
\begin{equation}
u= \sqrt{r^2{\cal Z}_{*}+\sigma^2}  \quad \& \quad \tan \hat{\alpha}=\frac{\sigma}{\sqrt{{\cal Z}_{*}}r}  \quad {\rm with} \quad 0<\hat{\alpha}<\frac{\pi}{2} \ .
\end{equation}
The function ${\cal Z}$ as well as the dilaton generally depend on both coordinates,
$u$ and $\hat{\alpha}$, but in the UV  limit of large $u$ the $\hat{\alpha}$ dependence disappears. Actually,  their expressions when $u\to\infty$ are:
\begin{equation}
{\cal Z} \rightarrow  {\cal Z}_{*} \quad \& \quad e^{-2\Phi} \rightarrow 
\frac{{\cal Z}_{*}^{3/4}}{2\sqrt{2} } \left(\frac{R}{u}\right)^{3/2} \ .
\label{UV-functions}
\end{equation}
Using these values in the metric ansatz we end up with the following expression:
\begin{eqnarray} \label{UV}
ds^2_{UV} & \approx &{ 2\sqrt{2}\over  {\cal Z}_{*}^{3/4}} 
\left(\frac{u}{R}\right)^{3/2}
\left[dx^2_{1,2}+{\cal Z}_{*}
R^2 d\tilde{\Omega}_2^2\right] + \frac{1}{2\sqrt{2} \,{\cal Z}_{*}^{1/4}}
\left(\frac{R}{u}\right)^{3/2}du^2 + \nonumber \\
&&  + \frac{1}{2\sqrt{2} }\, { R^{3/2}\over{\cal Z}_{*}^{1/4}} \,
u^{1/2}\left[d\hat{\alpha}^2 + \cos^2 \hat\alpha \,d\Omega_2^2+\sin^2\hat{\alpha}\,\,
 (d\psi+\cos \tilde{\theta}\, d\tilde{\phi})^2\right] \ ,
\end{eqnarray}
where $d\tilde{\Omega}_2^2\equiv d\tilde{\theta}^2+\sin^2\tilde{\theta}\,d\tilde{\phi}^2$ is the line element of the $(\tilde\theta, \tilde\phi)$ two-sphere. In order to interpret the meaning of the results just found, let us recall that, given a background of type IIA theory such as the one we are considering, one can generate a solution of eleven-dimensional supergravity by uplifting the
metric by means of the standard  formula:
\beq
ds_{11}^2\,=\, e^{-{2\over 3}\Phi}\,ds^2_{10}\,+\,e^{{4\over 3}\Phi}\,(dz)^2\,\,,
\label{ds-11d}
\eeq
where $z$ is the eleventh M-theory coordinate. We shall apply (\ref{ds-11d}) to the ten-dimensional UV metric and dilaton written in eqs. (\ref{UV}) and (\ref{UV-functions}). After changing the radial variable $u$ by a new coordinate $y$, defined as:
\beq
y^2\,=\,{2R\over \sqrt{{\cal Z}_*}}\,\,u\,\,,
\eeq
the resulting eleven-dimensional UV metric becomes:
\bear
&&ds_{11}^2\,\approx\,{y^2\over R^2}\,
 \,\Big[\,dx^2_{1,3}\,+\,
{\cal Z}_*\,R^2\,d\tilde \Omega_2^2\,\Big]\,+\,
R^2\Big({dy\over y}\Big)^2\,+\,\rc\rc
&&\qquad\qquad
\,+\,{R^2\over 4}\,\left[d\hat{\alpha}^2 + \cos^2 \hat\alpha \,d\Omega_2^2+\sin^2\hat{\alpha}\,\,(d\psi+\cos \tilde{\theta}\, d\tilde{\phi})^2\right] \,\,,
\label{AdS7-S4}
\eear
where $dx^2_{1,3}\,=\,dx^2_{1,2}+dz^2$. From (\ref{AdS7-S4}) we conclude that the uplifted metric is of the form $AdS_7\times S^4$, with the $AdS_7$ having two of its directions compactified in a two-sphere and with the $S^4$ being  fibered over this $S^2$, Notice also that the radius of the  $AdS_7$ is just $R$, whereas the $S^4$ has radius $R/2$. These results are, of course, consistent with the origin of the solution \cite{Maldacena:2000mw}, as coming from M5 wrapped on a two-cycle.

%%%%%%%%%%%%%%%%%%%%%%%%%%%%%%%%%%%%%%%%%%%%%%%%%%%%%%%%%%%%%%%%%%%%%

\section{Addition of flavor branes}
\label{flavoredbackground}

In this section we will start exploring the possibility of finding the dual of the ${\cal N}=4$ 3d gauge theory with matter hypermultiplets in the fundamental representation of the gauge group. We will achieve this by adding flavor branes to the setup of section 
\ref{unflavored}.  These flavor branes will be D4-branes extended along the three Minkowski directions $x^0, x^1$ and $x^2$ as well as
the $\psi$ and $\sigma$ directions  of the Calabi-Yau. At the same time they will be  located at particular fixed values of the $S^2$ sphere and of the transverse $\mathbb{R}^{3}$, as represented in the following array:
\begin{center}
\begin{tabular}{|c|c|c|c|c|c|c|c|c|c|c|}
\multicolumn{4}{c}{ }&
\multicolumn{4}{c}{$\overbrace{\phantom{\qquad\qquad\qquad}}^{\text{CY}_2}$}\\
\hline
&\multicolumn{3}{|c|}{$\mathbb{R}^{1,2}$}
&\multicolumn{2}{|c|}{$S^2$}
&\multicolumn{2}{|c|}{$N_2$}
&\multicolumn{3}{|c|}{$\mathbb{R}^{3}$}\\
\hline
$N_c$\,\,\,D$4$ (color) &$-$&$-$&$-$&$\bigcirc$&$\bigcirc$&$\cdot$&$\cdot$&$\cdot$&$\cdot$&$\cdot$\\
\hline
$N_f$\,\,\,D$4$ (flavor) &$-$&$-$&$-$&$\cdot$&$\cdot$&$-$&$-$&$\cdot$&$\cdot$&$\cdot$\\
\hline
\end{tabular}
\label{44flavored-array}
\end{center}
It is worth pointing out that, in the above setup, the flavor branes wrap a non-compact direction of the internal Calabi-Yau. This is, actually, a standard requirement which one should demand of these kinds of setup in order to convert, in the appropriate decoupling limit,  the gauge symmetry living on the worldvolume of the  flavor brane into a global (flavor) symmetry. 

To describe more precisely the embedding of the flavor brane let us choose the following system of worldvolume coordinates:
\beq
\xi^{\alpha}\,=\,(x^0, x^1, x^2, \sigma, \psi)\,\,.
\label{wv-coordinates}
\eeq
Then, the embedding is determined by the condition that all other remaining ten-dimensional coordinates are constant. By analyzing the kappa symmetry of the worldvolume action of the flavor brane we will explicitly check in appendix \ref{embeddings} that the embedding just described preserves the same supersymmetry as the unflavored setup.
Moreover, the position in $r$ of the flavor brane, $r_Q$,  represents the distance between the two sets of branes and has a well-defined meaning in the gauge theory dual. Indeed, $r_Q$ is related to the mass $m_Q$ of the matter hypermultiplet by means of the following relation:
\beq
m_Q={r_Q\over 2\pi\alpha'}\,\,.
\label{mQ}
\eeq

In the so-called quenched approximation the effects of the quark loops in the field theory observables are neglected. This approximation is well justified when $N_f\ll N_c$ and corresponds, on the gravity side, to treating the flavor branes as probes and neglecting their influence on the metric. Later in this paper we will make use of this approximation to study several aspects of the gauge theory, such as the meson spectrum and the quark-antiquark potentials. However, in the remainder of this section we will analyze, in a certain approximation,  how the backreaction of the flavor branes modifies the solution described in section \ref{unflavored}.

\subsection{Including the backreaction}
\label{backreaction}

Let us study the backreaction of the flavor branes on the background in the case in which the number of flavors $N_f$ is large and of the same order as the number of colors $N_c$. From the field theory point of view this limit was considered a long time ago by Veneziano \cite{Veneziano:1976wm}. Here we will follow the approach pioneered in ref. \cite{Casero:2006pt}, which is based on the observation that, when $N_f\to\infty$, one can homogeneously distribute the $N_f$ flavor branes in their transverse directions (for a clear discussion on the validity of the DBI+WZ action for a large number of smeared branes, see section 7 of \cite{HoyosBadajoz:2008fw}). Notice that, when the branes are embedded as explained around (\ref{wv-coordinates}), they preserve the same supersymmetries independent of their position in the transverse space. Actually, we will consider a distribution of branes with a fixed value $r_Q$ of the $r$ coordinate and smeared along the angular coordinates $(\theta, \phi)$ and 
 $(\tilde\theta, \tilde\phi)$. In order to figure out how this smearing is implemented, let us recall that the action for a stack of $N_f$ D4-branes is given by the sum
of the  DBI and WZ terms:
\begin{equation}
S_{flavor}\,=\,-T_4\,\sum_{N_f}\,\int_{{\cal M}_5}\,d^{5}\xi\,e^{-\Phi}\,
\sqrt{-\det\hat G_5}\,+\,T_4\,\sum_{N_f}\,\int_{{\cal M}_5}\,{\hat C}_5\,\,,
\label{flavor-DBI-WZ}
\end{equation}
with $\hat G_5$ being the induced metric on the worldvolume and
${\hat C}_5$ the pullback of the RR four-form potential to ${\cal M}_5$.
The smearing procedure amounts to promoting the infinite sum
appearing in the action to a ten-dimensional integral. For the WZ part we have:
\beq
\sum_{N_f}\,\int_{{\cal M}_5}\,{\hat C}_5 \quad \rightarrow \quad \int_{{\cal M}_{10}}\,
\Omega\wedge C_5\,\,,
\eeq
where $\Omega$ is a five-form proportional to the volume  form of the transverse space, namely:
\begin{equation}
\Omega\,=\,{N_f\over 16\pi^2}\,\,\delta(r-r_Q)\,dr \, \wedge \, \omega_2 \, \wedge \tilde{\omega}_2 \, ,
\label{Omega}
\end{equation}
with $\omega_2 $ being given by (\ref{omega2}) and $\tilde\omega_2=\sin\tilde\theta d\tilde\theta\wedge d\tilde\phi$. The smearing form $\Omega$ is normalized as:
\beq
\int \Omega\,=\,N_f \, .
\eeq
Similarly, to obtain the smeared version of  the DBI part of the flavor brane action we perform the substitution:
\begin{equation}
\sum_{N_f}\,\int_{{\cal M}_5}\,d^{5}\xi\,\,e^{-\Phi}\,\sqrt{-\det\hat G_5}\,
\rightarrow\,\int_{{\cal M}_{10}}\,d^{10} x\,\,e^{-\Phi}\,
\sqrt{- \det G}\,\big|\,\Omega\,\big|\,\,,
\end{equation}
where $\big|\,\Omega\,\big|$ is the modulus of $\Omega$:
\beq
\Big|\,\Omega\,\Big|\,=\,\sqrt{{1\over 5!}\,
\Omega_{M_1\cdots M_5}\,
\Omega_{N_1\cdots N_5}\,\,
\prod_{k=1}^{5}\,G^{M_k N_k}}\,\,.
\label{modulusOmega}
\eeq
Therefore,  the smeared DBI+WZ action of the flavor branes is:
\beq
S_{flavor}\,=\,-T_4\,
\int_{{\cal M}_{10}}\,d^{10} x\,\,e^{-\Phi}\,
\sqrt{-\det G}\,\big|\,\Omega\,\big|\,+\,
T_4\,\,\int_{{\cal M}_{10}}\Omega\wedge C_5\,\,.
\label{smearedDBI}
\eeq
By inspecting the WZ term in the action (\ref{smearedDBI}) one readily concludes that the flavor brane acts as a source for the RR six-form $F_6=dC_5$ which, in turn, induces a violation of the Bianchi identity of $F_4={}^*F_6$. Actually, one can prove that this modified Bianchi identity becomes:
%
%\footnote{In general, the relevant part of an action for a form $F_{(n)}=d\,A_{(n-1)}$ and the equation of motion for this
%form are
%
%\begin{equation}
%-\frac{1}{2n!}\,\,\int \sqrt{|g|}\, F^2 +\int G \wedge A \quad \Rightarrow \quad d \star F \, = \, sign(g)(-1)^{D-n+1} G \, .
%\end{equation}
%
%In our case, the relevant part of
%the action, in the string frame, and the equation of motion read
%
%\begin{equation}
%-\frac{1}{2\kappa_{10}^2}\frac{1}{2\cdot 6!}\int \sqrt{-g} \, F_{6}^2
%\,+\, T_4 \int \Omega \wedge C_{5}  \quad \Rightarrow \quad
%\frac{1}{2\kappa_{10}^2}\,\, d\star F_{6}\,=\,T_4 \,\, \Omega \, .
%\end{equation}
%
%Taking into account $F_{4}=\star\,\, F_{6}$ we arrive at
%\eqref{newBianchi}.}
%
\begin{equation}
dF_4\,=\,2\,\kappa_{10}^2 \, T_4\,\Omega\,=\,\frac{N_f}{2 N_c}\frac{R^3}{8}\,\delta(r-r_Q)
\,dr\wedge\, \omega_2 \, \wedge \, \tilde{\omega}_2 \, .
\label{newBianchi}
\end{equation}
where, in the last step, we have used that $2\kappa_{10}^2\,=\,(2\pi)^7\,g_s^2\,(\alpha')^4$ and $T_4=1/(2\pi)^4 g_s (\alpha')^{{5\over 2}}$ and, thus,  $2\kappa_{10}^2\,=\,\pi^2\,R^3/N_c$. 

Let us now formulate a new ansatz for this backreacted flavored setup. First of all, we will adopt the same ansatz (\ref{metric-ansatz}) for the ten-dimensional metric. However, as is clear from (\ref{newBianchi}), we should change the ansatz for $F_4$ in order to reproduce the modified Bianchi identity. Actually,  the natural ansatz both satisfying \eqref{newBianchi} and generalizing
\eqref{F4-form} is:
\begin{equation} \label{F4-flavor}
F_4\,=\,-\,(\dot{g}\,d\sigma\,+\,g' \, dr)\,\wedge \, \omega_2 \, \wedge \, (d\psi+\cos\tilde{\theta} \,d\tilde{\phi})\,+\,\left(\,g\,+\,\frac{N_f}{2 N_c}\frac{R^3}{8}\,\,\Theta(r-r_Q)
\, \right)\,\,\tilde{\omega}_2 \, \wedge \, \omega_2 \, ,
\end{equation}
where $\Theta$ is the Heaviside step function. 
Proceeding as in the unflavored case and substituting the new ansatz  \eqref{F4-flavor}
for $F_4$ into the equations for the supersymmetric variations of dilatino and
gravitino we have:
\begin{eqnarray}
 \label{flavored-BPSsystem}
&&\left[\,g\,+\,{N_f \over 2\,N_c}\,{R^3 \over
8}\,\Theta(r-r_Q)\,\right]\,=-\ \,R^{2}\,r^2\,{\cal Z}'  \ , \quad \quad
e^{-4\Phi}\,\sigma\,=\,R^{2}\,{\cal Z\dot{Z}} \ , \rc
\\
&& g'\,=\,-4\,\sigma\,r^2\,e^{-4\phi}\dot{\Phi} \ , \rc\rc
&&\dot g\,=\,\,-\sigma\,R^{-2}\,{\cal
Z}^{-2}\,e^{-4\Phi}\,\left[\,g\,+\,{N_f \over 2\,N_c}\,{R^3
\over 8}\,\Theta(r-r_Q)\,\right]\,+\,4\,\sigma\,r^2\,{\cal
Z}^{-1}\,e^{-4\phi}\,\Phi' \, .  \nonumber
\end{eqnarray}
The set of projections on the Killing spinors needed to arrive at
\eqref{flavored-BPSsystem} is the same as in the unflavored case. Thus, the flavored solutions preserve also the same eight supersymmetries as the unflavored ones. Moreover, as in (\ref{no-flavor-BPSsystem}), the last equation in the system (\ref{flavored-BPSsystem}) is not independent of the others. One can also verify (see appendix \ref{BPSeqs}) that the equations of motion of $F_4$ and the metric (including the contribution of the DBI action to Einstein's equations) are satisfied if the system (\ref{flavored-BPSsystem}) holds. 
The analogue of the  PDE (\ref{PDE-z-no-flavored})  for the flavored case is:
\be r^2\,{\cal Z}\,\left(\,\dot{{\cal Z}}\,-\,\sigma\, \ddot{{\cal
Z}}\,\right)\,=\,r\,\sigma\,\left(\,r\,\dot{{\cal Z}}^2\,+\, r\,{\cal
Z}''\,+\, 2 {\cal Z}'\,\right)\,+\sigma\,{N_f \over 2\,N_c}\,{R^3
\over 8}\,\delta(r-r_Q). \label{PDE-z-flavored} \ee
As in the unflavored case, we can integrate the function ${\cal Z}$ for
$\sigma=0$. Indeed, it follows from \eqref{flavored-BPSsystem}
that $g$ is independent of $r$ when the variable $\sigma$
vanishes. If we call $g_0$ this constant value of $g$ then,
for $r >r_Q$, we have:
\be {\cal Z}'(r, 0)\,=\,-\,{1 \over R^2}\,\left[\,g_0\,+\,{N_f
\over 2\,N_c}\,{R^3 \over 8}\,\right]\,\,{1\over r^2} \, , 
\qquad\qquad (r >r_Q)\,\,,
\ee
and after using for $g_0$ the same value as in the unflavored case, we get:
\be \label{zprime-sigmazero-flavor}  {\cal Z}'(r, 0)\,=\,{R \over 8\,r^2}
\left[1\,-\,{N_f\over 2\,N_c}\,\right]\,\,,
\qquad\qquad (r >r_Q) \, .
\ee
When $r<r_Q$, by simply putting $N_f=0$ on the right-hand side of \eqref{zprime-sigmazero-flavor}, we recover the unflavored result. Integrating \eqref{zprime-sigmazero-flavor} and imposing continuity for the solution along
$r=r_Q$ we have:
\beq
{\cal Z}(r, 0)\,=\,{\cal Z}_{*}\,- \,{R \over 8\,r_Q }
\,{N_f\over 2\,N_c}\,\Theta(r-r_Q)\,-\,{R\over 8\,r}
\left[\,1\,-\,{N_f\over 2\,N_c}\,\Theta(r-r_Q)\,\right]
\, ,\label{z-sigma0-flavor}
\eeq
\begin{figure}[ht]
\begin{center}
\includegraphics[scale=1]{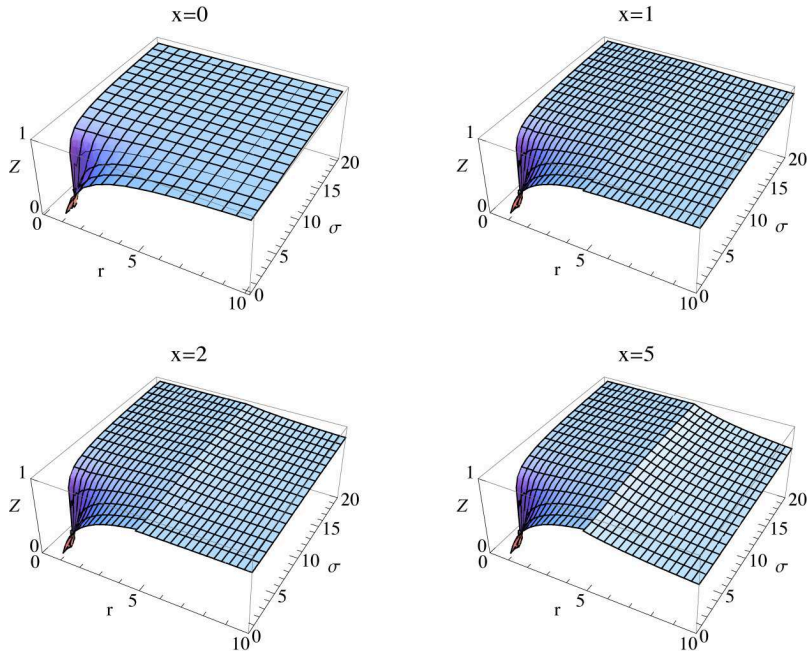}\label{Zs}
\caption{\em 
Solutions for $\cal Z$ for different values of $x=\frac{N_f}{N_c}$ as a function of $r$ and $\sigma$. These curves correspond to $\kappa={\cal Z}_*=1$ and $r_Q=5$.
}
\end{center}
\end{figure}
where the constant ${\cal Z}_{*}$ is the same as in \eqref{z-sigma0}. We have not been able to obtain analytically  the solution of the flavored system (\ref{flavored-BPSsystem}) for arbitrary values of $r$ and $\sigma$. Instead, one can integrate numerically this system of equations. Since the system  (\ref{flavored-BPSsystem}) reduces to the unflavored one written in (\ref{no-flavor-BPSsystem}), we can assume that the solution of (\ref{flavored-BPSsystem}) reduces to the one written in (\ref{GammaZ})-(\ref{g-Phi}) for $r<r_Q$, while at $r=r_Q$ the functions ${\cal Z}$, $g$ and $\Phi$  are continuous and 
${\cal Z}'$ has a discontinuity independent of $\sigma$ and given by:
\beq
{\cal Z}'(r_Q+\epsilon, \sigma)\,-\,{\cal Z}'(r_Q-\epsilon, \sigma)\,=\,-{R\over 8 r_Q^2}\,
{N_f\over 2N_c}\,\,.
\eeq
Using the continuity and change in derivative of $Z$ at $r=r_Q$ and \ref{z-sigma0-flavor} as the boundary conditions, we are able to numerically integrate the equation of motion for $Z$ up to a finite value of $r$ after which the solution becomes highly oscillatory and unstable.
The result of the numerical integration for ${\cal Z}(r,\sigma)$ is shown in figure \ref{Zs} for several values of $N_c$ and $N_f$.  It is quite evident from figure \ref{Zs} that ${\cal Z}$ has a wedge shape at the position $r=r_Q$ of the flavor branes. This means that   there will be a curvature singularity at this point, which is actually needed to match a similar term in the Einstein equations coming from the energy-momentum tensor of the flavor brane sources.  As in the unflavored case, the geometry is also singular in the IR due to the enhan\c{c}on phenomenon. Outside these regions the metric is regular and one can verify that the components of the Ricci tensor depend on $N_c$ and $N_f$ through the combination $N_c-{Nf\over 2}$, which scales as $N_c$ in the Veneziano limit 
$N_c, N_f\to\infty$ with $N_c/N_f$ fixed.  Thus, outside the location of the source, the region in which the supergravity approximation is valid is not modified by the backreaction of the flavor branes, as happened in the backgrounds of refs. \cite{Casero:2006pt,Paredes:2006wb,Arean:2008az}.

\section{Connection with gauge theory}
\label{gaugetheory}

The ${\cal N}=4$ three-dimensional gauge theories can be regarded as the reduction to three dimensions of the ${\cal N}=2$ gauge theories in four dimensions \cite{Seiberg:1996nz}. The field content of the different  3d  supermultiplets can be easily obtained by dimensional reduction  of the one  corresponding  to ${\cal N}=2, d=4$. Let us consider, for example, the vector multiplet which, in four dimensions is composed by a vector field, a Dirac spinor and a complex scalar. By reducing to three dimensions one gets one vector, two Dirac fermions and three scalars. Similarly, one can verify that the  ${\cal N}=2, d=3$ hypermultiplet contains two Dirac fermions and four complex scalars. Given this field content it is straightforward to find the one loop running coupling constant  
$g_{YM}(\mu)$ of the theory (see, for example, the appendix C of \cite{Divecchia}). Indeed, let us consider a $SU(N_c)$ gauge theory with $N_f$ matter hypermultiplets in the fundamental representation. Then, one can show that:
\be {1\over g_{YM}^2(\mu)}\,=\, {1\over
g^2_{YM}}\,\Big[\,1\,-\,{g^2_{YM}\,N_c \over
4\pi\mu}\,\Big(\,1-{N_f\over 2\,N_c}\,\Big) \,\,\Big] \ ,
\label{gYM-QFT} 
\ee
where $\mu$ is the energy scale. The one-loop result displayed in eq. (\ref{gYM-QFT}) is, actually, exact in perturbation theory. Notice that  \eqref{gYM-QFT} shows that the ${\cal N}=4$ theory has negative beta
function when $N_f<2N_c$, while for $N_f>2N_c$  the beta function
changes its sign and becomes positive. In the borderline case
$N_f=2N_c$ the one-loop beta function vanishes and the coupling does
not run anymore  in perturbation theory. In the next subsection we will show that our gravity solutions match perfectly the behavior (\ref{gYM-QFT}), both in the unflavored and backreacted flavored cases.

Besides the perturbative result just reviewed, the ${\cal N}=4$  3d theories have a very rich non-perturbative structure. Indeed, the Coulomb branch of vacua of these theories is a hyperk\"ahler manifold, which is isomorphic to the moduli space of  three-dimensional monopole solutions of a different gauge theory \cite{Seiberg:1996nz, Hanany:1996ie}. Moreover, the Higgs branch   is also a  hyperk\"ahler manifold. Furthermore, these theories display the phenomenon of mirror symmetry \cite{Intriligator:1996ex,deBoer:1996mp}, which is a duality between two different ${\cal N}=4$ 3d gauge theories which exchanges the Higgs and Coulomb branches, as well as the Fayet-Iliopoulos and mass terms. When these  ${\cal N}=4$ theories are realized in the type IIB
string theory as in \cite{Hanany:1996ie}, the mirror symmetry is just a manifestation of the underlying S-duality of the type IIB string theory (see ref. \cite{GiveonKarch} for a review).

\subsection{Probe calculation}

In order to extract information about the gauge theory living  on the D4-branes, we will study, following ref. \cite{Divecchia},  the dynamics of a color  D4-brane probe wrapping an $S^2$ and moving under the influence of the metric and RR form of the background. The action of such a probe will be:
\beq \label{actionD4}
S\,=\,-T_4\,\int d^5\xi\,e^{-\Phi}\,\,\sqrt{-\det (\,\hat G_5\,+\,2\pi\alpha'\,F)}\,+\,
T_4\,\int \left( \hat C_5+ 2\pi \alpha' \, \hat{C}_3 \wedge F \right)\,\,,
 \eeq
where $\xi^a$ ($a=0,\cdots, 4$) is the set of worldvolume coordinates along which the color D4-brane is extended, $F$ is the field strength for the worldvolume gauge field and the hat over $G_5$, $C_5$ and $C_3$ denotes the pullback over the worldvolume of the D4-brane. In particular, let us evaluate the action (\ref{actionD4}) in the case in which the ten-dimensional metric is of the form (\ref{metric-ansatz}). Let us choose the worldvolume coordinates  as $\xi^a=(x^0, x^1, x^2, \tilde\theta, \tilde\phi)$ and let us assume that the brane in embedded in such a way that the remaining ten-dimensional coordinates are constant and the worldvolume gauge field $F$  is zero. Then, the determinant of the induced metric for such a configuration takes the form:
\beq
e^{-\Phi}\,
\sqrt{-\det \hat G_5}\,=\,R^2\,{\cal Z}\,e^{4\Phi}\,\sin\tilde{\theta}\,\,
\sqrt{1\,+\,\sigma^2\,{\cot^2\tilde{\theta}\over R^2\, {\cal Z}^2\,e^{4\Phi}}\,}\,\, .
\label{induced-metric}
\eeq
For the configuration we are considering, the only  non-vanishing contribution to the WZ part in (\ref{actionD4}) is the term containing $C_5$. The expression of this five-form potential for our solutions has been evaluated in appendix \ref{BPSeqs} (eq. \eqref{C5}). The corresponding pullback to the D4-brane worldvolume is:
\beq
\hat C_5\,=\,R^2 \,{\cal Z}\,e^{4\Phi}\, \,\sin\tilde{\theta}\,\,dx^0\wedge dx^1\wedge dx^2\wedge d\tilde{\theta} \wedge d\tilde{\phi}
\,\, .
\label{induced-C5}
\eeq
Let us now substitute (\ref{induced-metric}) and (\ref{induced-C5}) into (\ref{actionD4}).  The result is just minus  the static potential between the stack of $N_c$ color branes and the probe, namely:
\beq
S_{pot}\,=\,-T_4\,\int d^3 x\,d\tilde\theta d\tilde\phi\,\,R^2\,{\cal Z}\,e^{4\Phi}\, 
\sin\tilde\theta\,\,
\Big[\, \sqrt{1\,+\,\sigma^2\,{\cot^2\tilde{\theta}\over R^2\, {\cal Z}^2\,e^{4\Phi}}}
\,-\,1\,\Big]\,\,.
\label{potD4}
\eeq
Notice that the right-hand side of (\ref{potD4}) only vanishes when $\sigma=0$, which should be interpreted as the point of the Calabi-Yau in which one can place a color brane without breaking supersymmetry. Actually, one can check this statement directly by studying the implementation of kappa symmetry for the different D4-brane embeddings (see subsection \ref{Higgsbranch} and appendix \ref{embeddings}). Let us thus assume that our probe brane is located at $\sigma=0$ and that we switch on a worldvolume gauge field $F_{\mu\nu}$ whose only non-vanishing components are those along the Minkowski directions $x^{\mu}$. We will expand the DBI lagrangian density (integrated over the angular directions $\tilde\theta$ and $\tilde\phi$) up to quadratic order in the gauge field $F_{\mu\nu}$. The result of this expansion can be parameterized as:
\beq
 \int\,d\tilde\theta d\tilde\phi\, {\cal L}_{DBI}\Big |_{quadratic}\,=\,
\,-{1\over 2\,g_{YM}^2(\mu)}\,
  tr[ F_{\mu\nu}F^{\mu\nu}]\,\,,
  \label{quadraticDBI}
 \eeq
where $g_{YM}(\mu)$ is, by definition, the Yang-Mills coupling at the renormalization scale $\mu$.  Actually, by performing explicitly the calculation, one gets:
\be {1\over g^2_{YM}(\mu)}\,=\,{R^2 \over 2\,\pi g_s
(\alpha')^{1/2}}\,\,{\cal Z}(r,\sigma=0)\,\,.
 \label{gYM-gravity} \ee
Clearly, the variable $r$ should be related to the energy scale $\mu$ of the field theory.
The natural radius-energy relation is given by:
\be r\,=\,2\pi\,\alpha'\,\mu\, , 
\label{radius-energy}
\ee
which we will assume to hold in the following.
Let us now suppose that $r>r_Q$ or, equivalently, that $\mu>m_Q$, where $m_Q$ has been defined in (\ref{mQ}). After substituting in \eqref{gYM-gravity} the value of ${\cal Z}(r, \sigma=0)$ as given by \eqref{z-sigma0-flavor}
for the general flavored solution we have:
\be 
{1\over g_{YM}^2(\mu)}\,=\,{{\cal Z}_{*}\,R^2 \over
2\,\pi \, g_s \, (\alpha')^{1/2}} \,-\, \frac{N_f}{8\,\pi\,m_Q}\,-\,{N_c \over
4\,\pi\,\mu}\,\left(\,1\,-\,\frac{N_f}{2\,N_c}\right)\,\,.
\label{gYM-mu}
\ee
Moreover, by defining the bare UV YM coupling as:
\be
 \label{UV-YM-coupling} \frac{1}{g^2_{YM}}\,=\,{{\cal Z}_{*}\,R^2 \over
2\,\pi \, g_s \, (\alpha')^{1/2}} \,-\,\frac{N_f}{8\,\pi\,m_Q}
\, , \ee
we get that eq. (\ref{gYM-mu}) matches perfectly the field theory expression (\ref{gYM-QFT}).  Moreover, by allowing the brane probe to move in the transverse flat space, and by looking at the action of the transverse scalar fields obtained by expanding the DBI+WZ action, one can obtain the metric of the moduli space in the Coulomb branch. After dualizing the worldvolume gauge field one can check that the moduli space is indeed hyperk\"ahler and, actually, its metric has the form of the Taub-NUT space \cite{Divecchia}.

\subsection{Higgs branch}
\label{Higgsbranch}

The Higgs branch of the ${\cal N}=4$ gauge theory is a phase in which the quark hypermultiplets acquire a non-vanishing expectation value. On the field theory side one can study the theory in the Higgs branch by turning on an extra Fayet-Ilioupoulos coupling in the lagrangian. The realization of this mechanism in   a brane setup is well-known \cite{Hanany:1996ie} (see \cite{GiveonKarch} for a review). Indeed, as argued in ref. \cite{Hanany:1996ie}, one should reconnect the color and flavor branes in a supersymmetric way. In our holographic setup we should look for D4-brane embeddings which are compatible with all the supersymmetries of the gravity solution and such that they can be interpreted as representing a recombination of color and flavor branes (see \cite{Arean:2008az, CEGK, Arean:2007nh} for a similar analysis in other brane setups). Recall that both types of D4-branes are extended along different directions of the Calabi-Yau cone. Indeed, the color branes are extended along $(\tilde\theta, \tilde\phi)$ at $\sigma=0$ while the flavor branes extend along $(\sigma, \psi)$ at fixed angles $(\tilde\theta, \tilde\phi)$. In order to find a configuration interpolating between these two situations it is natural to use the same system of worldvolume coordinates as in (\ref{wv-coordinates}) and look for an embedding such that $\tilde\theta$ and $\tilde\phi$ are no longer constant. Instead, they will depend on the other coordinates of the $CY_2$, namely:
\beq
\tilde\theta\,=\,\tilde\theta(\sigma,\psi)\,\,,
\qquad\qquad
\tilde\phi\,=\,\tilde\phi(\sigma,\psi)\,\,.
\label{embedding-ansatz}
\eeq
In order to determine the D4-brane embeddings of the form (\ref{embedding-ansatz}) which preserve the supersymmetries of the background, one has to study the kappa symmetry of the brane probe. This analysis is performed in detail in appendix \ref{embeddings}. The final result found in this appendix can be nicely recast in terms of the following two complex coordinates of the $CY_2$: 
\beq
\zeta_1\,\equiv\,\tan \Big( {\tilde\theta\over 2}\Big)\,e^{i\tilde\phi}\,\,,\qquad\qquad
\zeta_2\,\equiv\, \sigma \sin \tilde\theta\,e^{-i\psi}\,\,.
\label{def-zetas}
\eeq
It turns out that any holomorphic embedding of the type $\zeta_1=f(\zeta_2)$, with $f$ arbitrary,  solves the kappa symmetry equations and, thus, preserves the supersymmetry of the background. In order to make contact with the field theory analysis, it is rather natural to restrict ourselves to embeddings characterized by a polynomial equation of the type:
\beq
\zeta_1^{p_1}\,\,\zeta_2^{p_2}\,=\,C\,\,,
\label{polynomial-embedding}
\eeq
with $C$ being a (complex) constant and the exponents $p_1$ and  $p_2$ are constant integers. Notice that, in the Coulomb branch, the embedding of the color branes correspond to eq. (\ref{polynomial-embedding}) with $p_1=0$, $p_2=1$ (and $C=0$), while that of the flavor brane is given by $p_1=1$, $p_2=0$. It is thus natural to think that the Higgs branch embedding we are seeking is the one obtained by taking $p_1=1$, $p_2=1$ in (\ref{polynomial-embedding}). Notice that, in this case, the functions (\ref{embedding-ansatz}) are just given by:
\beq
\sin^2\Big({\tilde\theta\over 2}\Big)\,=\,{\sigma_*\over \sigma}\,\,,
\qquad\qquad
\tilde\phi\,=\,\psi+\tilde\psi_0\,\,,
\label{tildetheta-tildepsi-p1}
\eeq
where $\sigma_*$ and $\tilde\psi_0$ are constants (related to $C$ in (\ref{polynomial-embedding})). Notice that  $\sigma_*$ is just the minimal value of the coordinate $\sigma$ (which occurs for $\tilde\theta=\pi$) which should correspond, in the field theory side, to the Higgs VEV.

\section{Meson spectrum}
\label{mesons}
In this section we analyze the  mass spectrum of the mesonic excitations for the backgrounds of sections \ref{unflavored} and \ref{flavoredbackground}.  In order to address this question we will add a flavor D4-brane probe to these backgrounds and we will study the normalizable fluctuations of its worldvolume fields. This analysis will lead us to a problem of the Sturm-Liouville type with a discrete set of eigenfunctions and eigenvalues. 

Let us consider a flavor D4-brane in a geometry of the form (\ref{metric-ansatz}). We will choose the system (\ref{wv-coordinates}) of worldvolume coordinates and we will consider an embedding in which the radial coordinate $r$ and the angular coordinates $\theta$, $\phi$, $\tilde\theta$ and $\tilde\phi$ are constants. Let us denote by $r_q$ the constant value of $r$ for this configuration ($r_q$ is related to the mass $m_q$ of the external quarks as in (\ref{mQ}), \ie\ $m_q=r_q/2\pi\alpha'$).  The induced metric for such a configuration is:
\beq
{\cal G}_{ab}\,d\xi^a\,d\xi^b\,=\,
e^{2\Phi(r_q, \sigma)}\,
\,dx^2_{1,2}\,+\,
{e^{-2\Phi(r_q, \sigma)}
\over{\cal  Z}(r_q, \sigma)}\,
\Big[\, d\sigma^2\,+\,\sigma^2\,(d\psi)^2\,\Big]\,\,.
\label{induced-metric-fluct}
\eeq
Let us now perturb this static configuration by deforming it in the radial direction as:
\beq
r\,=\, r_q\,+\,\hat r(x^{\mu}, \sigma, \psi)\,\,,
\label{r-fluct}
\eeq
where the fluctuation $\hat r$ is small. In all the calculations of this section we will take $r_q$ such that the $r=r_q$ surface does not enter the enhan\c{c}on region and, thus, $\sigma=0$ is the minimal value of the coordinate $\sigma$ in the worldvolume. This can always be achieved by taking $r_q$ sufficiently large.

Of course, the perturbation (\ref{r-fluct}) is not the most general one. However, one can check that, at quadratic order, $\hat r$ does not mix with other fluctuations and, therefore, it can be studied separately. By expanding the corresponding DBI+WZ action \footnote{One should take into account that, due to the different signs in the two projections in (\ref{projections}), the WZ coupling of the RR potential $C_5$ for the flavor brane must be opposite to the one appearing in (\ref{actionD4}). } one can verify that, up to quadratic terms, one gets that the lagrangian density for $\hat r$ is:
\beq
{\cal L}\,=\,-{T_4\over 2}\,e^{-3\Phi}\,
\sqrt{-\det {\cal G}}\,{\cal G}^{ab}\,\,
\partial_a\,\hat r\,\partial_b\,\hat r\,=\,
-{T_4\over 2}\,\Big[\,{\sigma \over e^{4\Phi} {\cal Z}}\,\,
(\partial_{x^{\mu}}\hat r)^2\,+\,
\sigma (\partial_{\sigma}\hat r)^2\,+\,{1\over \sigma}\,
(\partial_{\psi}\hat r)^2\,\Big]\,\,,
\label{lagrangian-fluct}
\eeq
where $\Phi$ and ${\cal Z}$ should be understood as functions of $\sigma$ at $r=r_q$.
The equation derived from the lagrangian (\ref{lagrangian-fluct}) is given by:
\beq
\partial_{\sigma}\,\big[\,\sigma\partial_{\sigma}\hat r\,\big]\,+\,
{\sigma \over e^{4\Phi} {\cal Z}}\,\,
\partial^2_{x^{\mu}}\,\hat r\,+\,{1\over \sigma}\,\partial_{\psi}^2\,
\hat r\,=\,0\,\,.
\label{hatr-eom}
\eeq
To find the solutions of this equation, let us separate variables as:
\beq
\hat r\,=\,\chi(\sigma)\,e^{ikx}\,e^{il\psi}\,\,,
\label{hatr-separated}
\eeq
where $l$ is an integer (which can be taken to be non-negative without loss of generality) and $k$ is a momentum along the Minkowski  directions $x^{\mu}$. Let us also define $M^2$ as $M^2\equiv -k^2$, where the square is computed with the flat Minkowski metric in $2+1$ dimensions. Plugging (\ref{hatr-separated}) into (\ref{hatr-eom}), we arrive at the following equation for $\chi(\sigma)$:
\beq
\partial_{\sigma}\,\big[\,\sigma\partial_{\sigma}\chi\,\big]\,+\,\Big[\,
{\sigma \over e^{4\Phi} {\cal Z}}\,\,M^2\,
\,-\,{l^2\over \sigma}\,\Big]\chi\,=\,0\,\,,
\label{separated-fluct}
\eeq
Interestingly,  by means of a suitable change of variables, the fluctuation equation (\ref{separated-fluct}) can be written as a Schr\" odinger equation. Indeed, let us define the variable $y$ as follows:
\beq
e^{y}\,=\,\sigma\,\,.
\eeq
Notice that $y\in(-\infty, +\infty)$.  In terms of $y$, the equation (\ref{separated-fluct}) of the fluctuations can be written as:
\beq
{d^2\chi\over dy^2}\,-\,V(y)\,\chi\,=\,0\,\,,
\label{Sch}
\eeq
where the potential $V(y)$ is given by:
\beq
V(y)\,=\,l^2\,-\,M^2\,{e^{2y}\over e^{4\Phi} {\cal Z}}\,\,.
\label{potential-Sch}
\eeq
From  the reformulation of eqs.  (\ref{Sch}) and (\ref{potential-Sch}) of the fluctuation equation, one can easily obtain the asymptotic value of  $\chi$ when $y\to \pm\infty$. Indeed, from the behavior of $\Phi(r_q,\sigma)$ and ${\cal Z}(r_q,\sigma)$ when $\sigma\to 0,\infty$ one easily gets that $V\to l^2$ when $y\to\pm\infty$.  It follows that, in these asymptotic regions, the two independent solutions of (\ref{Sch}) when $\l\not=0$  are just 
$\chi\sim e^{\pm l y}=\sigma^{\pm l}$ (when $l=0$ these solutions behave as 
$\chi={\rm constant}, \log\sigma$ when $\sigma\to 0, \infty$). The normalizable solutions, which can be identified with mesonic excitations, are those that are regular as $\sigma=0,\infty$. These solutions only exist for some discrete set of values of the mass $M$, which can be determined numerically by means of the shooting technique (see below). 

The previous analysis only applies to the particular fluctuation $\hat r$ in the radial direction of the transverse ${\mathbb R}^3$. However, one can check that the same equation (\ref{separated-fluct}) describes the fluctuations of the other two ${\mathbb R}^3$ coordinates $\tilde\theta$ and $\tilde\phi$. Furthermore, in order to study the fluctuations of the coordinates $\tilde\theta$ and $\tilde\phi$ that determine the position of the flavor brane in the $CY_2$, let us write:
\beq
\tilde\theta\,=\,\tilde\theta_0\,+\,\hat\theta (x^{\mu}, \psi, \sigma)\,\,,
\qquad
\tilde\phi\,=\,\tilde\phi_0\,+\,\hat\phi (x^{\mu}, \psi, \sigma)\,\,,
\label{fluct-in-CY2}
\eeq
where $\tilde\theta_0$ and $\tilde\phi_0$ are the constant unperturbed values of the angles $\tilde\theta$ and $\tilde\phi$. By plugging (\ref{fluct-in-CY2}) into the DBI+WZ action and expanding the result up to quadratic order in the fluctuations one gets a lagrangian in which $\hat\theta$ and $\hat\phi$ are coupled. It turns out that, as happened in refs. \cite{Paredes:2006wb} and \cite{Arean:2008az}, they can be easily decoupled by defining new fields $\chi_+(\sigma)$ and
$\chi_-(\sigma)$ as follows:
\bear
&&\hat\theta\,=\,{1\over 2}\,\Big(\,\chi_+(\sigma)\,+\,\chi_-(\sigma)\,\Big)\,\sin\theta_0\,
e^{ikx}\,\sin (l\psi)\,\,,\rc\rc
&&\hat\phi\,=\,{1\over 2}\,\Big(\,\chi_+(\sigma)\,-\,\chi_-(\sigma)\,\Big)\,
e^{ikx}\,\cos (l\psi)\,\,,
\eear
where $l$ is a non-negative integer. One can verify that, indeed, $\chi_+(\sigma)$ and
$\chi_-(\sigma)$ satisfy two different decoupled equations. Actually, it is possible to perform a further redefinition to the functions $\chi_{\pm}$ in such a way that the new functions satisfy the scalar fluctuation equation (\ref{separated-fluct}). Similarly, one can show, as in ref. \cite{Paredes:2006wb}, that the fluctuations of the worldvolume gauge field also satisfy (\ref{separated-fluct}).  These facts imply that, up to finite shifts in  the quantum numbers, all fluctuation equations lead to the same set of eigenvalues, as expected from supersymmetry. Therefore, one can concentrate on studying the basic equation (\ref{separated-fluct}). This is what we will do in the next two subsections.

\subsection{Quenched mesons in the Coulomb branch}

\begin{figure}[ht]
\begin{center}
\includegraphics[scale=1.7]{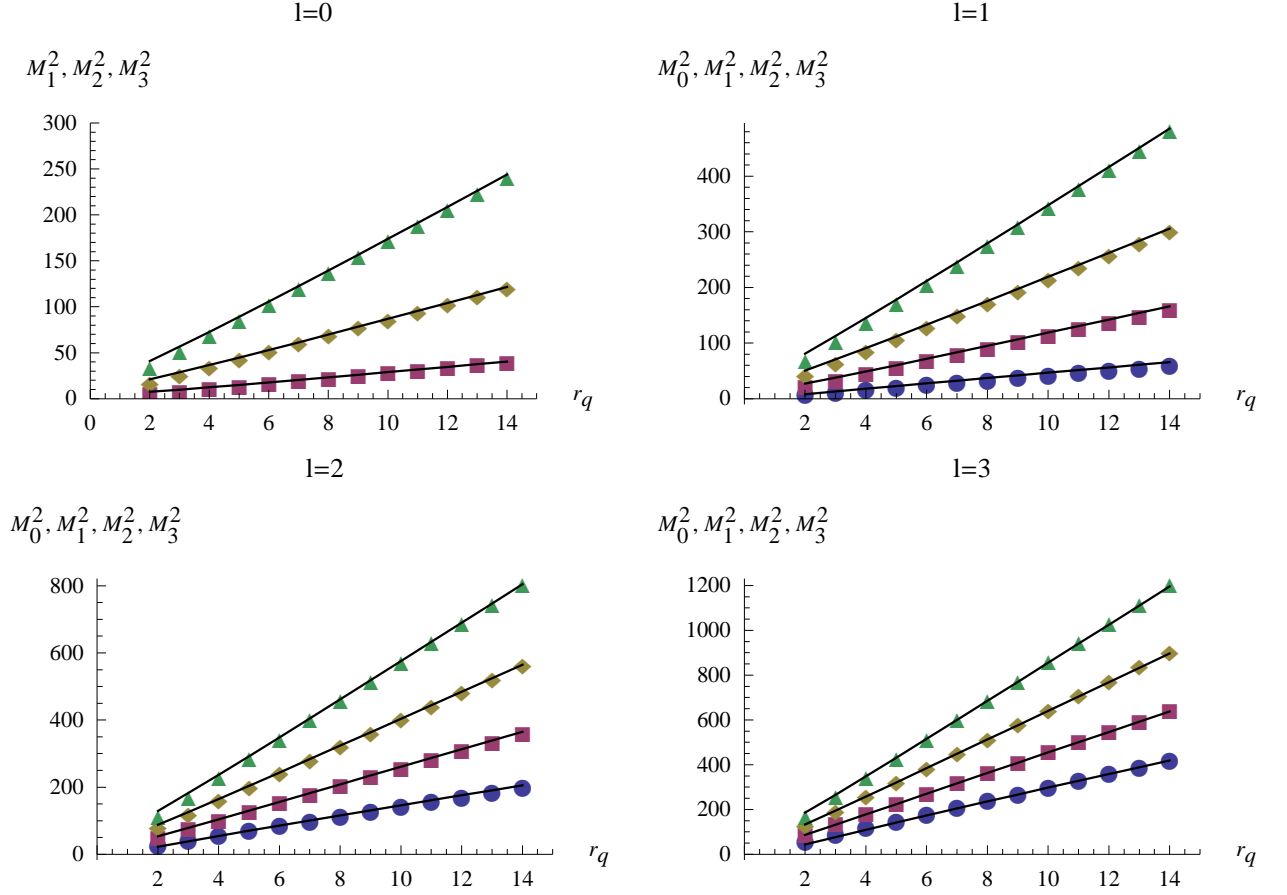}
\caption{\em \label{Coulombspectra}
Mass eigenvalues from the improved WKB formula  (eq. \ref{MWKB-improved}) (dots) and from the numerical calculation (solid line). The numerical calculation has been performed with the unflavored background with $\kappa={\cal Z}_*=1$.   }
\end{center}
\end{figure}

Let us  specialize the previous analysis to the case in which the background geometry is the unflavored one of section \ref{unflavored} and we add a flavor D4-brane probe. In this case the function ${\cal Z}$ and the dilaton $\Phi$ are given by eqs. (\ref{Zimplicit}) and (\ref{g-Phi}) respectively.  The numerical results obtained by means of the shooting technique are shown in figure \ref{Coulombspectra}. In general, for given $r_q$ and $l$, one obtains a tower of discrete normalizable states labelled by a principal quantum number $n$. It is interesting to notice that, for fixed $n$ and $l$, the masses grow with $r_q$  as $M^2\sim r_q$. 

One can get a rather accurate estimate of the mass spectrum by applying the semiclassical WKB method to the  Schr\" odinger problem (\ref{Sch}) (see refs. \cite{Minahan:1998tm}-\cite{Arean:2006pk}). By applying this method to our particular case, we get the following mass formula:
\beq
M^2_{WKB}\,=\,{\pi^2\over\big[ \zeta(r_q)\big]^2}\,\,(n+1)(n+3l)\,\,,
\label{MWKB}
\eeq
where $n$ is a non-negative integer ($n\ge 0$ when $l\not=0$ and $n\ge 1$ for $l=0$). In eq. (\ref{MWKB})  $\zeta(r_q)$ is the following integral:
\beq
\zeta(r_q)\,=\,\int_0^{+\infty}\,\,{d\sigma\over
\sqrt{e^{4\Phi(r_q, \sigma)}\,{\cal Z}(r_q, \sigma)}}\,\,.
\label{zeta-WKB}
\eeq
Notice that   $\zeta(r_q)$ determines the mass gap of the mesonic spectrum. One can get an approximate expression for $\zeta(r_q)$ by using inside the integral on the right-hand side of (\ref{zeta-WKB}) our asymptotic expressions (\ref{UV-Z}) and (\ref{UV-H}). By doing so one arrives at the following analytic expression for  $\zeta(r_q)$:
\beq
\zeta(r_q)\,\approx\,\sqrt{{R^3\over 8}}\,\,{\cal Z}_*^{{1\over 4}}\,
\int_0^{+\infty}\,
{d\sigma\over (\sigma^2\,+\,{\cal Z}_*\,r^2_q)^{{3\over 4}}}\,=\,
\sqrt{{\pi^3\,R^3\over 16}}\,{1\over \Big[\Gamma\big({3\over 4}\big)\Big]^2}\,\,
{1\over \sqrt{r_q}}\,\,.
\label{zeta-approx}
\eeq
It is interesting to stress that, within this approximation, $\zeta(r_q)$  is independent of ${\cal Z}_*$. Actually, by using the value of $R$ written in (\ref{R}), as well as the relation between $r_q$ and the quark mass $m_q$ (see (\ref{mQ})), one can write the following WKB formula for the meson masses:
\beq
M^2_{WKB}\,=\,
{4\,\Big[\Gamma\big({3\over 4}\big)\Big]^4\,\,m_q\over \pi\,g_s\,N_c\,\sqrt{\alpha'}}\,\,
\Big(n+1\Big)\,\Big(n+3l\Big)\,\,.
\label{MWKB-approx}
\eeq
This formula reproduces rather well the numerical values of the masses for $l=0,1$ and, actually, captures accurately the meson mass gap and its dependence on the quark mass $m_q$. However, the degeneracies observed in the numerical results when one varies  the quantum numbers $n$ and $l$ are not reproduced by (\ref{MWKB-approx}). Nevertheless we found that the numerical results are fully recovered if (\ref{MWKB-approx}) is changed to:
\beq
M^2\,=\,
{4\,\Big[\Gamma\big({3\over 4}\Big)\Big]^4\,\,m_q\over  \pi\,g_s\,N_c\,\sqrt{\alpha'}}\,\,
\Big(n+l\Big)\,\Big(n+2l+1\Big)\,\,.
\label{MWKB-improved}
\eeq
The comparison between the predictions of (\ref{MWKB-improved}) and the numerical values of the masses is shown in figure \ref{Coulombspectra}. It follows from the inspection of  this figure that  (\ref{MWKB-improved}) provides a very good fit of the masses found by the shooting technique.

\subsection{Unquenched mesons in the Coulomb branch}

We can perform the analysis of the meson spectrum by looking at the fluctuations of a flavor brane probe in the background given by the backreacted flavor branes. This is equivalent to studying the fluctuations of a single quenched flavor in the presence of $N_f$ unquenched flavors. In the large $N_f$ limit this should be an accurate approximation. What is interesting to note here is that because the functions ${\cal Z}$ and $\Phi$ are themselves continuous in $r$ and are being taken at the value $r=r_q$, the eigenvalues for a flavor probe at this position will be exactly the same as that for a probe without the backreacted flavor.

Thus, 
by inspection we see from the form of the equations of motion for the probe brane fluctuations that the spectrum will be identical in the flavored and unflavored case if the probe brane is placed at the same position as the backreacting branes. The difference between the quenched and unquenched case will only be felt if we introduce a finite distance between the probe and backreacting branes, corresponding to Higgsing a $U(1)$ of the $U(N_f)$ flavor group.

\begin{figure}[ht]
\begin{center}
\includegraphics[scale=1.65]{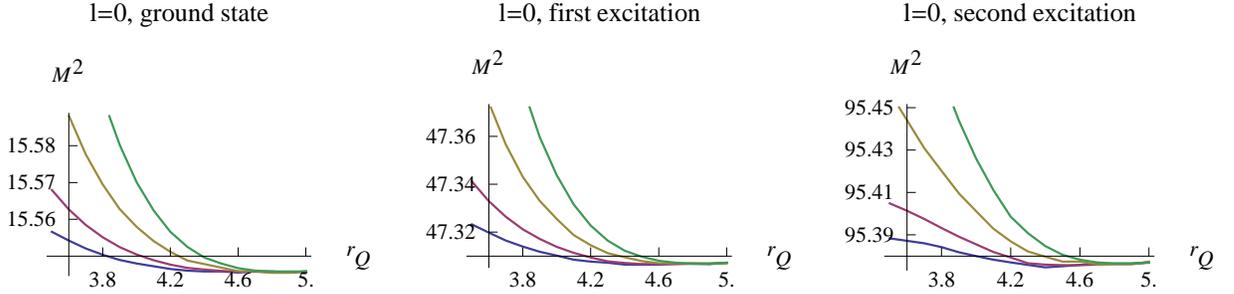}
\caption{\em \label{flavouredcoulomb}
Variation of the meson spectrum calculated on a probe brane at $r_q=5$ with backreacting flavor branes at varying positions $r_Q$. The four lines on each plot correspond to $N_f/N_c=1$ (flattest line), $N_f/N_c=2$, $N_f/N_c=5$ and $N_f/N_c=10$ (least flat line). It is clear to see that each of these return to the quenched case in the case where the backreacting branes lie close to the probe brane. All calculations have been done for $\kappa={\cal Z}_*=1$.
}
\end{center}
\end{figure}

The case where the probe and backreacting branes are separated is considerably more complicated, computationally, because in order to calculate the meson spectrum it is vital to have a numerically stable geometry. The calculation of the function ${\cal Z}$ from the partial differential equation is inherently an unstable calculation and therefore the calculation of the spectrum is difficult. We have however managed to calculate the spectrum in a narrow range of parameter space where the probe and backreacting branes are not very far apart.

In figure \ref{flavouredcoulomb} we study the effects of placing the backreacting branes at a variable position, given by $r_Q$ while always keeping the position of the probe brane at $r_q=5$. On each graph the lines correspond to $N_f/N_c=10, N_f/N_c=5, N_f/N_c=2$ and $N_f/N_c=1$. Clearly the $N_f/N_c=1$ line is the flattest in each case and the $N_f/N_c=10$ line has the most variation, as expected as the geometry is most altered with a higher ratio of flavor branes to color branes. We have checked that the curves in figure \ref{flavouredcoulomb} can be fitted to an expression of the type 
$M^2=a\,+\,N_f\,b\,\big(\,{r_Q\over r_q}\,-\,1\,\big)^4$, where $a$ and $b$ are coefficients that are independent of $N_f$.

\subsection{Mesons in the Higgs branch}
 We will now consider the fluctuations around a non-trivial embedding of the type studied in subsection \ref{Higgsbranch} (eq. (\ref{polynomial-embedding})). Recall that we argued in this subsection that these embeddings correspond to configuration in which  both types of D4-branes (color and flavor) are recombined, realizing the Higgs branch in our brane setup.  More concretely, we will concentrate on studying the embedding 
 (\ref{polynomial-embedding}) for $p_1=p_2=1$, which can be rewritten as in (\ref{tildetheta-tildepsi-p1}). We will continue to use (\ref{wv-coordinates}) as our system of worldvolume coordinates but now $\tilde\theta$ and $\tilde\phi$ will no longer be constant but given by the functions displayed in (\ref{tildetheta-tildepsi-p1}). For simplicity we will restrict ourselves to studying the fluctuations of the radial coordinate $r$ around a fixed value $r_q$ (see eq. (\ref{r-fluct})). After separating variables as in (\ref{hatr-separated}) the fluctuations equation can be written in the Schr\" odinger form (\ref{Sch}), if we introduce a new variable $y$ as:
\beq
e^y\,=\,\sigma\,-\,\sigma_*\,\,,\qquad\qquad
-\infty\,<\,y\,<+\infty\,\,,
\eeq
where $\sigma_*$ is the minimal value of the coordinate $\sigma$ for the embedding (see eq. (\ref{tildetheta-tildepsi-p1})). Recall that $\sigma_*$ parameterizes the Higgs VEV. The Schr\" odinger potential $V$ is now given by:
\beq
V\,=\,{l^2\over 4}\,-\,M^2\,\Bigg[\,
{e^{2y}\over e^{4\Phi}\,{\cal Z}}\,+\,\sigma_*\,R^2\,{e^y\,{\cal Z}\over
(e^y\,+\,\sigma_*)^2}\,\Big]\,\,.
\label{potential-Higgs}
\eeq
Note here that in the limit $l\rightarrow 2l$ and $\sigma_\star\rightarrow 0$ we recover the Coulomb branch potential. However, one can check that  this limit is not continuous and the last term in (\ref{potential-Higgs}) means that the Coulomb spectrum is never recovered taking this limit in the Higgs branch.

In order to have an idea of the mass spectrum associated to the potential (\ref{potential-Higgs}), let us apply the WKB method. The mass gap is now determined by an integral similar to (\ref{zeta-WKB}), which is now given by:
\beq
\zeta(r_q, \sigma_*)\,=\,\int_{\sigma_*}^{\infty}\,d\sigma\,\sqrt{
{1\over e^{4\Phi(r_q, \sigma)}\,
{\cal Z}(r_q, \sigma)}\,+\,\sigma_*\,R^2\,
{{\cal Z}(r_q, \sigma)
\over \sigma^2 \,(\sigma-\sigma_*)}}\,\,.
\label{zeta-higgs}
\eeq
When $\sigma_*$ is not very small the second term inside the square root in (\ref{zeta-higgs}) dominates the integral. Actually, by applying the same approximations that led to (\ref{zeta-approx}), one arrives at the following approximate expression for $\zeta(r_q, \sigma_*)$:
\beq
\zeta(r_q, \sigma_*)\,\approx\,\pi\,R\,\sqrt{{\cal Z}_*}\,\,.
\label{z-approx-higgs}
\eeq
Notice that $\zeta$ in (\ref{z-approx-higgs}) is independent of $r_q$ and $\sigma_*$ and only depends on the asymptotic value of ${\cal Z}_*$. Actually, after improving the WKB formula as in (\ref{MWKB-improved}), we arrive at the approximate formula for the mass levels in the Higgs branch:
\beq
M^2_{Higgs}\,\approx\,
{1\over R^2\,{\cal Z}_*}\,\,
\big(n+l\big)\,\big(n+l+1\big)\,\,,
\eeq
which reproduces rather well the numerical values.

\section{Wilson loops and the $q\bar{q}$ potential}
\label{Wilsonloops}

In order to investigate the unflavored background more, we now
turn to another direction of a more phenomenological nature, the Wilson loop
operator. On the gravity side, the Wilson loop expectation value is calculated by minimizing the Nambu--Goto action for a fundamental string stretching into the dual
supergravity background, whose endpoints are constrained to lie on
the two sides of the Wilson loop. Below, we first briefly review
the procedure for calculating Wilson loops in supergravity. We will discover that the qualitative behavior of this observable critically depends on whether the constant 
$\kappa$, appearing in the function (\ref{GammaZ}),  is smaller or larger than $1/16$. We will discuss these two cases separately in two different subsections. 

%%%%%%%%%%%%%%%%%%%%%%%%%%%%%%%%%%%%%%%%%%%%%%%%%

\subsection{General formalism}

As stated above, the calculation of a Wilson loop in the gravity
approach amounts to extremizing the Nambu--Goto action for
a string propagating in the dual geometry whose endpoints trace
the loop. To describe such configuration in our setup let us choose the time $t$ and a Minkowski coordinate $x$ as worldvolume coordinates of the string and let us consider 
the trajectory with:
\be
\label{track} r\,=\,r(x)\ ,\quad \sigma \,=\, 0\ ,\quad {\rm rest}\, =\, {\rm constant} \ .
\ee
Then, the induced metric on the string worldvolume can be easily found from (\ref{metric-ansatz}), namely:
\begin{equation} \label{reduced-ansatz}
e^{2\Phi (r,0)}\,\big[\,-dt^2\,+\,(1+e^{-4\Phi (r,0)}\,\,r^{\prime\, 2})\,dx^2\,\big]\,\,,
\end{equation}
with $\Phi (r,0)$ being given by (\ref{dilaton-at-sigma0}), which we can rewrite in terms of the enhan\c{c}on radius $r_e$ defined in (\ref{re}) as:
\begin{equation} \label{dilaton-sigma-zero}
e^{4\Phi (r,0)}\,=\,\frac{1}{r_e R^2 {\cal Z}_{*}}
\, \frac{r^2\left(\,r^2-2r r_e + 16\,\kappa\, r_e^2\,\right)}{r-r_e} \,.
\end{equation}
The Nambu-Goto action for the configuration (\ref{track}) is:
\be
\label{NG}
S = \frac {1}{2\pi} \int dt\, dx \,\,\sqrt{e^{4\Phi (r,0)}\, +\,  r^{\prime 2}} \,,
\ee
where we have taken $\alpha'=1$ and  the prime denotes a derivative with respect to $x$. Since the action does not explicitly depend on $x$, the system has a first integral $r_0$,  which can be identified with the turning
point of the solution. Solving the corresponding first-order
equation for $x$ in terms of $r$ we find that the linear
separation of the quark and antiquark is:
\be \label{wilson-length}
L= 2  \int^\infty_{r_0} dr \,\,{e^{2\Phi (r_0,0)}\over e^{2\Phi (r,0)}
 \sqrt{e^{4\Phi (r,0)}\,-\,e^{4\Phi (r_0,0)}}}\ .
\ee
The energy of the configuration can be obtained from the action \eqref{NG}. Subtracting the self-energy contribution, we obtain
\be
\label{wilson-energy}
E = {1 \over \pi} \int^\infty_{r_0} dr \,\,
{e^{2\Phi (r_0,0)}\over 
 \sqrt{e^{4\Phi (r,0)}\,-\,e^{4\Phi (r_0,0)}}}
- {1\over \pi} \int^{\infty}_{r_{\rm min}} dr \ ,
\ee
where $r_{\rm min}$ is the minimum value of $r$ allowed by the
geometry. In the specific cases, we are supposed to solve for the
auxiliary parameter $r_0$ in terms of the separation length
$L$. Since this cannot be done explicitly in practice, one regards \eqref{wilson-length} as a parametric equation
for $L$ in terms of the integration constant $r_0$. Combining it
with \eqref{wilson-energy} for $E$, one can then determine the behavior of the potential energy of the configuration in terms of the quark-antiquark separation. Using \eqref{wilson-length} and \eqref{wilson-energy} we arrive at the following relation which will prove useful in the following:
\beq
{d E_{q\bar q}\over  d L}  =  {e^{2\Phi (r_0,0)}\over  2\pi } \, .
\label{first-deriv-energy}
\eeq

The minimum value of $r$ allowed by the geometry depends on $\kappa$.  Indeed, as argued at the end of subsection \ref{unflavored-integration}, the numerator in the right-hand side of (\ref{dilaton-sigma-zero}) has real roots only when $\kappa\le 1/16$.  In this case $r_{min}\,=\,r_H$, where $r_H$ has been written in (\ref{rH}).
Conversely, when $\kappa>1/16$ the minimum value of $r$ is $r_{min}\,=\,r_e$.

\begin{figure} 
\begin{center}
\includegraphics[height=5.5cm]{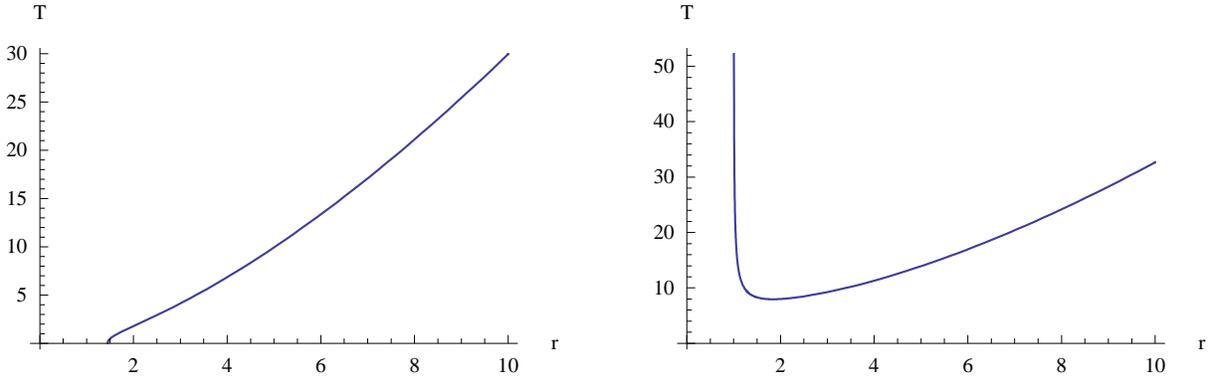}\\
\end{center}
\caption{Effective tension of the string as a function of $r$. On the left is the case with $\kappa\leq 1/16$ and on the right with  $\kappa>1/16$.}
\label{tension-representation}
\end{figure}

Following the analysis in \cite{HoyosBadajoz:2008fw}, we can determine the  qualitative behavior of the system simply by looking on the asymptotic expansion of the dilaton in \eqref{dilaton-sigma-zero}.  Utilizing their formalism we write the lagrangian as:
\beq
{\cal L} = T(r) \sqrt{1+ Y(r)\, r'^2} \, , 
\eeq
where $T(r)$ is the effective tension of the brane depending on the dilaton. In our background this is given by:
\begin{equation}
T(r)\,=\,{1 \over Y(r)^{1/2}}\,=\,e^{2\Phi(r,0)} \, .
\end{equation}
When $\kappa\leq 1/16$  both the tension of the string and $Y$ behave in the IR with a power law, $T={1 \over Y^{1/2}}\sim (r-r_{min})^{1/2}$. Following the reasoning of ref. \cite{HoyosBadajoz:2008fw}, due to the value of the exponent of $Y$, the string will not reach an infinite length  in the $x$ direction when it explores the far IR of the dual theory. As we will see in the next subsection this is verified both by our numerical and  analytical results.

When $\kappa> 1/16$ the string feels an effective IR wall at  the point $r=r_*$ where the tension is minimized, so $T'(r_*)=0$  and $T''(r_*)>0$ (see figure \ref{tension-representation}).  The actual value of $r_*$ can be obtained analytically by finding the roots of a cubic polynomial. These roots depend on $\kappa$ and $r_e$ and one can check that $r_*\to 2r_e$ when $\kappa$ is large. 
Since the background at the point $r=r_*$  is smooth and the functions of the metric 
do not vanish, they admit a Taylor expansion as:
\begin{eqnarray}
&& T=T_* +T_2 (r-r_*)^2+\dots, \quad  T_2 >0, \nonumber \\
&& Y=Y_*+Y_1 (r-r_*) +\dots. \, .
\end{eqnarray}
Following again the reasoning in \cite{HoyosBadajoz:2008fw}, when the tip of the string comes close  to the IR wall  the length will diverge. Again this is something we will verify in the next subsection.

%%%%%%%%%%%%%%%%%%%%%%%%%%%%%%%%%%%%%%%%%%%%%%%%%%%

\subsection{Case with $\kappa\leq 1/16$}

\begin{figure}
\begin{center}
\begin{tabular}{cc}
\includegraphics [height=6cm]{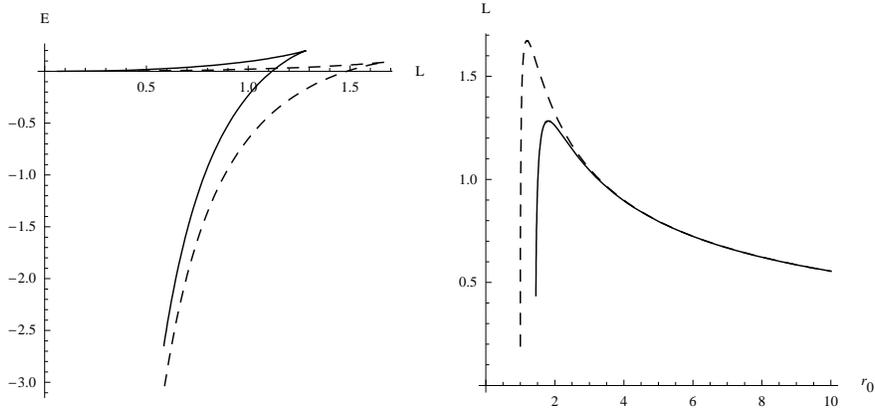}
\end{tabular}
\end{center}
\caption{$\kappa\leq 1/16$: Separation length between a quark and a antiquark  as a function of $r_0$ and energy as a function of the separation length. The dashed (solid) curve is the case for $\kappa=1/16$ ($\kappa=1/18$).}
\label{EL}
\end{figure}

 For large values of $r_0$ the potential is, as usual, Coulombic while their exists a maximal separation $\kappa\leq1/16$. At the values of $r_0$ where the separation is maximal, the energy is also maximal and always positive (see figure \ref{EL}). This means that there is a screening behavior because as the potential turns positive, a configuration of two separate strings is energetically favored and corresponds to a vanishing force between the charges. 
As $r_0\,\rightarrow \,r_{min}=r_H$ we find by studying  the behavior of the integrals in  eqs. (\ref{wilson-length}) and (\ref{wilson-energy}) in this case that both length and energy approach zero as:
\begin{equation}\label{wilson-length-k<16-approx}
L\,\backsimeq \,\sqrt{2}\,R\,\left(\frac{r_e {\cal Z}_{*}}{r_{H}}\right)^{1/2}\,\sqrt{\frac{r_0}{r_{H}}-1}\,\,
\ln\left[\frac{3\,r_{H}}{r_0-r_{H}}\right] \, ,
\end{equation}
and
\begin{equation}\label{wilson-energy-k<16-approx}
E\,\backsimeq \, \frac{r_0 - r_{H}}{2\pi}\,\ln\left[\frac{3\,r_{H}}{r_0-r_{H}}\right] \, .
\end{equation}

%%%%%%%%%%%%%%%%%%%%%%%%%%%%%%%%%%%%%%%%%%%%%%%%%%%

\subsection{Case with $\kappa > 1/16$}

In this case we follow the same steps as before but now we should have in mind that $r_{min}=r_e$. As usual, the dependence of the potential energy for small separations $L$
of the quark-antiquark (corresponding to large $r_0$) is Coulombic. In order to have an approximate expression for the separation length and the energy we have to move to the region of the parameter space where $\kappa$ is large. In this case the effective IR wall is located at  $r=r_*\backsimeq 2 r_e$ and the potential gives a linear confining behavior with:
\begin{equation}\label{length-confinement}
L\,\backsimeq \,\frac{{\cal Z}_{*}^{1/2}}{\sqrt{3+4\kappa}}\,\ln\left[\frac{6\,r_{e}}{r_0-2 r_{e}}\right] \,,
\end{equation}
and
\begin{equation}\label{Energy-confinement}
E\,\backsimeq \,\frac{4\,r_e}{\pi}\,\sqrt{\frac{\kappa}{3+4\kappa}}\,\ln\left[\frac{6\,r_{e}}{r_0-2 r_{e}}\right]\, \backsimeq \frac{4 r_e}{\pi}\,\sqrt{\frac{\kappa}{{\cal Z}_{*}}}\, L \, .
\end{equation}
\begin{figure}
\begin{center}
\includegraphics[height=7.7cm]{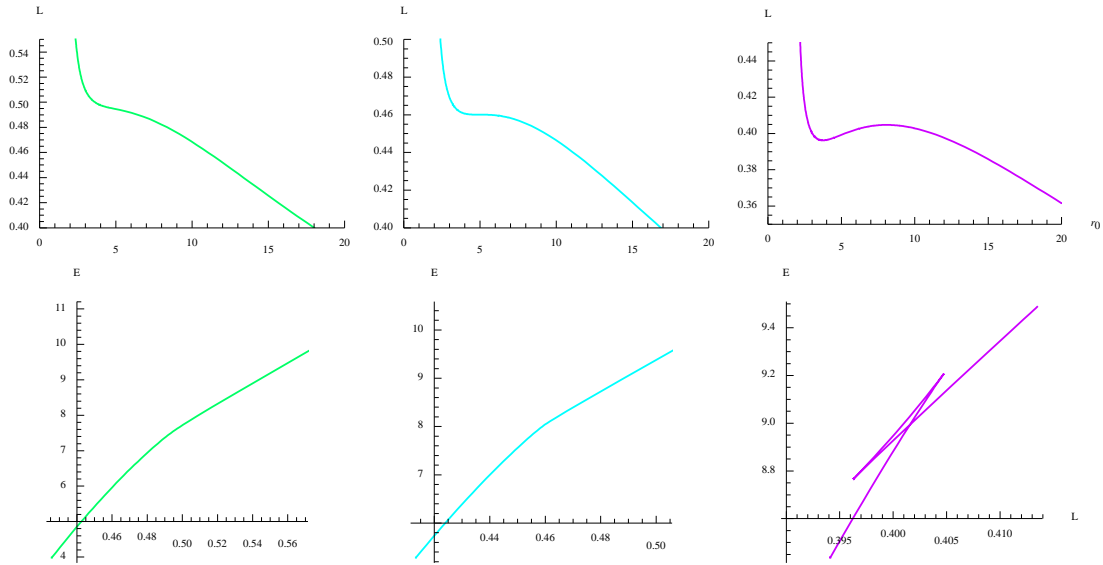}\\
\end{center}
\caption{$\kappa> 1/16$: Separation length between a quark and an antiquark  as a function of $r_0$ and energy as a function of the separation length. The left (right) curves correspond to the case in which $\kappa$ is below (above) the critical value and the central curves are  the case in which $\kappa$ has  exactly the critical value,  $\kappa_{\rm cr}\simeq 63.8$.}
\label{critical}
\end{figure}
For intermediate values of the separation length,
the behavior of the system depends crucially on the value of $\kappa$.
There is a critical value $\kappa_{\rm cr}\simeq 63.8$, such that for $\kappa>\kappa_{\rm cr}$ the behavior
of the length and energy curves resemble the Van der Waals isotherms for a statistical
system, with $r_0$, $L$ and $E$ corresponding to volume, pressure
and Gibbs potential respectively (see, for instance, \cite{Callen}).  In the following we will describe the situation referring to the corresponding plots in figure \ref{critical}.  In the
region below the critical point, $\kappa<\kappa_{\rm cr}$, the
energy is a single-valued function of the length, connecting a Coulombic with a linear confining phase.
For values above the critical point, $\kappa>\kappa_{\rm cr}$, the Coulombic phase for the energy at small distances is followed by a situation where the energy becomes a
triple-valued function of the length, with the physical state being the one of lowest energy. For large distances the energy returns to a single-valued
function of the length with approximately linear behavior. The self-intersection point in the energy curve indicates the presence of a first-order phase transition with order parameter $r_0$.  At this point we should note that exactly at the critical value, $\kappa_{\rm cr}\simeq 63.8$, the first order phase transition degenerates to a second order one. Then using purely thermodynamical arguments and the fact that not only the first but also the second derivative
of the length vanishes at the critical value $r_0^{\rm cr}$, we have:
\begin{equation} \label{length-critical}
L-L_{\rm cr}\sim (r_0-r_0^{\rm cr})^3\, ,
\end{equation}
which is in agreement with gravity calculations using  \eqref{wilson-length}. The critical behavior vanishes for
$\kappa<\kappa_{\rm cr}$. When $\kappa=\kappa_{\rm cr}$ so the corresponding critical exponent takes the classical value 3. Using \eqref{first-deriv-energy} and \eqref{length-critical} we can calculate the behavior of the energy close to the critical point:
\be \label{energy-critical}
E_{q\bar q} - E_{q\bar q}^{\rm cr}  \backsimeq\ 
{e^{2\Phi (r_0^{\rm cr},0)}\over 2 \pi}
(L-L_{\rm cr} ) \left(1- A |L-L_{\rm cr}|^{1/3}\right)\ ,
\ee
where $A$ is a constant. The above critical behavior is
similar to that found in \cite{Brandhuber:1999jr,Avramis:2007wb,Bigazzi:2008gd,Bigazzi:2008cc}. This expression is calculated numerically by tuning $\kappa$ very finely such that the first order phase transition, seen in $L(r_0)$, just disappears. In order to do this we must zoom into the region in the $L(r_0)$ plot where both $L'(r_0)$ and $L''(r_0)$ simultaneously vanish. Using this value of $\kappa$ we are then able to fit the parametric plot of $E$ and $L$ to very high accuracy and obtain the classical values to high numerical precision.

We close this section with the remark that the confining behavior appearing in the IR is not something expected from a gauge theory point of view.
For this reason, it would be interesting to examine the stability of the string trajectory used for calculating the quark--antiquark potential under small fluctuations,
through the tools obtained in \cite{Avramis:2006nv,Avramis:2007mv}.

\section{Discussion}
\label{conclusions}

In this paper we have studied the addition of flavor to the supergravity dual of three-dimensional gauge theories with eight supersymmetries. The unflavored background is constructed by wrapping D4-branes in a two cycle of a Calabi-Yau twofold. The corresponding flavor branes are also D4-branes that extend along the non-compact directions of the normal bundle of the cycle. We have shown that the addition of these flavor branes does not break any of the eight supersymmetries preserved by the unflavored background. We have studied this system both in the quenched and unquenched approaches and we have obtained the meson spectrum by analyzing the normalizable fluctuations of the probe flavor D4-branes. In the brane probe approximation we have also studied the Higgs branch of the theory, which is realized as a configuration in which the color and flavor D4-branes are recombined. 

We have succeeded in computing the backreaction of a large number of flavor branes by considering a continuous distribution of D4-branes smeared over their transverse angular directions.  The flavor branes provide a source term for the RR fields that induces a violation of the Bianchi identity for $F_4$. The ansatz for the backreacted background can be obtained by modifying the unflavored one to satisfy the new Bianchi identities.  By requiring that the flavored background preserves the same amount of supersymmetry as the unflavored solution, one arrives at a system of first-order BPS equations. We have checked that the BPS equations imply the Maxwell-Einstein equations with sources. The functions in the ansatz depend on the two radial variables
$\sigma$ and $r$ and the BPS equations  are a system of PDEs that must be integrated numerically. However, the solution for $\sigma=0$ can be found analytically and we have shown that it reproduces the running of the coupling of the Yang-Mills theory with flavors. 

The supergravity solutions we have studied  do not capture the rich non-perturbative structure of the corresponding field theory duals. This fact is related to the bad IR behavior of the solutions, which have a naked singularity in this region. This problem is similar to the one presented by other  similar backgrounds with the same amount of supersymmetry in four \cite{N2d4,Paredes:2006wb} and two dimensions \cite{Arean:2008az}.  As in these cases, one can argue that the singularity can be consistently screened by an enhan\c{c}on, which is the locus where the sources of the background become effectively tensionless and the geometry ends. The fact that the gravity solution reproduces the exact perturbative behavior and misses the non-perturbative effects seems to be due to the suppression of the latter in the 't Hooft large $N_c$ limit.

We have also studied the behavior of Wilson loops in the unflavored background, which critically depends on the value of the parameter $\kappa$ of the solution. While for $\kappa \le 1/16$ there is a maximal separation of the quark-antiquark pair, for 
$\kappa >1/16$ there is a transition from a Coulomb-like behavior at small separation to a linear potential for large separation.  Above a certain critical value of $\kappa=\kappa_{cr}$ the transition is discontinuous (first order) while exactly at the critical value of $\kappa$ it becomes second order. At this point we have estimated the critical exponents, which are given  by the classical mean field theory values. Actually, this critical behavior seems to be universal, at least for these type of models: we have found it in the 4d system of  \cite{N2d4}, as well as in the 2d system of  \cite{Arean:2008az}.  In both cases there is a constant which is the analogue of $\kappa$ and a critical value for this constant. Similar results have been found in \cite{Brandhuber:1999jr,Arean:2005ar,Avramis:2007wb, Bigazzi:2008gd,Bigazzi:2008zt,Bigazzi:2008cc} in other backgrounds.   Interestingly, the critical exponents in these systems are also given by the classical values. To properly interpret these results one should have a clean understanding of the meaning of the constant $\kappa$ (and of its  analogues in other models). The most natural interpretation is that $\kappa$ parametrizes the VEV of some operator that is switched on. However, more work is needed to confirm this interpretation  and to find the precise nature of the phase transition uncovered. We are working in this direction.

\section*{Acknowledgments}

We are grateful to Daniel Are\'an, Spyros Avramis, Francesco Bigazzi,  Aldo Cotrone, Carlos N\'u\~nez,  Angel Paredes  and Konstadinos Siampos for discussions and encouragement. This  work was supported in part by MEC and  FEDER  under grant
FPA2005-00188,  by the Spanish Consolider-Ingenio 2010 Programme CPAN (CSD2007-00042), by Xunta de Galicia (Conselleria de Educacion and grant PGIDIT06PXIB206185PR) and by  the EC Commission under  grant MRTN-CT-2004-005104.

\appendix

\section{BPS equations }
\label{BPSeqs}
Let us derive the BPS equations (\ref{flavored-BPSsystem}) of the general flavored background by imposing the preservation of eight supersymmetries.  For the type of background we are studying, the supersymmetry variations for the dilatino $\lambda$ and gravitino $\psi_M$  in the type IIA theory in the string frame are:
\begin{eqnarray} \label{susy-tranf}
&&\delta\lambda\,=\,{1\over 2}\,\left[\Gamma^{M}\,\partial_M\,\Phi -
\frac{1}{4\cdot 4!}\, e^{\Phi}F_{MNPQ}^{(4)}\,
\Gamma^{MNPQ}\right]\epsilon \ , \nonumber \\
&&\delta\psi_M\,=\,\left[\,\nabla_M\,-\,\frac{1}{8 \cdot 4!}
\,e^{\Phi}\,F_{NPQR}^{(4)}\,\Gamma^{NPQR}\,\Gamma_{M}\,\right]\,\epsilon \, ,
\end{eqnarray}
where $\epsilon$ is a ten-dimensional spinor. The Killing spinors of the background are those for which
$\delta\lambda=\delta\psi_M=0$.  They will be characterized by a set of algebraic projection conditions that can be expressed in terms of products of constant Dirac matrices with flat indices. In order to specify these conditions, let us choose the following vielbein basis: 
\begin{eqnarray}
&& e^{0,1,2}= e^{\Phi}dx^{0,1,2} \ ,  \quad e^3= e^{\Phi}{\cal Z}^{\frac{1}{2}}R\,d\tilde{\theta} \ , \quad
e^4 = e^{\Phi}{\cal Z}^{\frac{1}{2}}R\,\sin\tilde{\theta}d\tilde{\phi}\,\,,  \rc\rc
&& e^5= e^{-\Phi}{\cal Z}^{-\frac{1}{2}}d\sigma \ , \quad
e^6=e^{-\Phi}{\cal Z}^{-\frac{1}{2}}\sigma\left(d\psi+\cos\tilde{\theta}d\tilde{\phi}\right) \,, \label{frame}
\\ \rc
&&e^7= e^{-\Phi} dr \ , \quad e^8 = e^{-\Phi}r\,d\theta \ , \quad e^9= e^{-\Phi}r\sin\theta d\phi \,\,.
\nonumber 
\end{eqnarray}
It is useful to express the four-form (\ref{F4-flavor}) 
in terms of the vielbein basis (\ref{frame}). One has:
\bear
\label{F4-frame} 
&&F_4\,= \,-\,{e^{4\Phi} \over \sigma r^2}\,\,\sqrt{{\cal Z}}\,
\left(g'\,e^7 + \sqrt{{\cal Z}}\dot{g}\,e^5\,\right) \wedge
e^6\wedge e^8\wedge e^9\,+\,\rc\rc
&&\qquad\qquad\qquad
+\, {1 \over {\cal Z}r^2
R^2}\left[\,g\,+\,{N_f \over 2\,N_c}\,{R^3
\over 8}\,\Theta(r-r_Q)\,\right]\,e^3\wedge e^4\wedge e^8\wedge e^9\,\,,
 \eear
Then, the  projections satisfied by the Killing spinors are:
\be 
\Gamma_{01234}\,\Gamma_{11}\,\epsilon\,=\,\,\epsilon\,\,,\qquad\qquad
\Gamma_{01256}\,\Gamma_{11}\,\epsilon\,=\,- \epsilon \,\,, \label{projections} 
\ee
where the indices of the $\Gamma$'s refer to the frame basis (\ref{frame}). One can easily show that, 
after putting all these ingredients into the supersymmetry transformations for the gravitino and dilatino \eqref{susy-tranf}, we end up with the system  (\ref{flavored-BPSsystem}) of first-order BPS equations for the functions of our ansatz. Moreover,  the expression for the Killing spinors is:
\begin{equation} \label{killing spinor}
\epsilon\,=\,e^\frac{\Phi}{2}\,e^{-\frac{\psi}{2}\Gamma_{34}}\,e^{\frac{\theta}{2}
\Gamma_{78}}\,e^{\frac{\phi}{2}\Gamma_{89}}\,\,\eta\,\,,
\end{equation}
where $\eta$ is a constant spinor that satisfies the same
projections  (\ref{projections}) as $\epsilon$. Notice that these algebraic conditions imply that the system is $1/4$-supersymmetric, \ie\ that eight supersymmetries are preserved. 

Let us now show that the first-order BPS system (\ref{flavored-BPSsystem}) implies the second-order equations of motion of the different fields. First of all, let us start by checking the equation of motion of $F_4$, which is given by:
\begin{equation}
d({}^\star F_4) \,=\,0 \, ,
\label{F4-eom}
\end{equation}
One can prove that (\ref{F4-eom})  is equivalent to the following PDE:
\be \label{unflavor-PDEforF6}
\partial_{r}\,\Big[\,{{\cal Z} g 'e^{8\Phi} \over \sigma r^2 }\,\Big]\,+\,
\partial_{\sigma}\,\Big[\,{{\cal Z}^2 \dot g e^{8\Phi}\over \sigma r^2 }\,\Big]\,=\,
{\sigma \over R^4r^2 {\cal Z}^2}\,\,
\left[\,g\,+\,{N_f \over 2\,N_c}\,{R^3
\over 8}\,\Theta(r-r_Q)\,\right]
 \, .
\ee
Let us verify that equation \eqref{unflavor-PDEforF6}  is satisfied as a consequence of the system \eqref{flavored-BPSsystem}. We will check this fact by an explicit calculation. First, by using \eqref{flavored-BPSsystem}, we rewrite all terms appearing in \eqref{unflavor-PDEforF6} as:
\begin{eqnarray}
&&{{\cal Z} g' e^{8\Phi}\over \sigma r^2}\,=\,-\,\frac{\cal Z}{R^2}\,
\partial_{\sigma}\left(\frac{\sigma}{{\cal Z}\dot{{\cal Z}}}\right)\,\,,
\nonumber
\\
&&{{\cal Z}^2 \dot g\, e^{8\Phi}\over \sigma r^2}\,=\,\frac{1}{R^2}\,\partial_{r}\left(\frac{\sigma}{\dot{{\cal Z}}}\right)\,\,,
\nonumber
\\&& \frac{\,\sigma}{{\cal Z}^2\,r^2}
\,\,
\left[\,g\,+\,{N_f \over 2\,N_c}\,{R^3
\over 8}\,\Theta(r-r_Q)\,\right]
\,=\,R^2\,\partial_{r}\left(\frac{\sigma}{{\cal Z}}\right)\, .
\label{identities}
\end{eqnarray}
By plugging this result into \eqref{unflavor-PDEforF6} one can straightforwardly verify (\ref{F4-eom}).  Thus, $F_6={}^\star F_4$ should be represented as the derivative of a five-form, $dC_5$.  Indeed, by using again \eqref{flavored-BPSsystem} and (\ref{identities}) one can readily obtain an expression for this
five-form potential, namely:
\be \label{C5} C_5\,=\,dx^0\wedge dx^1\wedge dx^2 \wedge \left[\,
R^2\,{\cal Z}\,e^{4\Phi}\,
\,\tilde{\omega}_2\,-\,{\sigma \over {\cal
Z}}\,\,d\sigma\wedge (d\psi+\cos\tilde{\theta}
\,d\tilde{\phi})\,\right] \, .
\ee

In order to check that the Einstein and dilaton equations of motion
are satisfied in the flavored background, after using the BPS equations, we write them in the Einstein frame. In this frame we will use a one-form basis which is just the one in (\ref{frame}) conveniently rescaled with the exponential of the dilaton as:
\beq
E^{\bar M}\,=\,e^{-{\Phi\over 4}}\,\,e^{\bar M}\,\,.
\eeq
Notice that the smearing form $\Omega$ of (\ref{Omega}) can be written in this basis as:
\begin{equation}\label{Omega-Einstein}
\Omega \, = \, \frac{N_f}{16 \pi^2} \, \frac{e^{9\Phi/4}}{R^2}\,
\,\frac{\delta(r-r_Q)}{r^2 \, {\cal Z}} \,\, E^{3}
\wedge E^{4} \wedge E^{7} \wedge E^{8} \wedge E^{8}  \,.
\end{equation}
Moreover, the DBI action for the smeared flavor branes in the Einstein frame takes the form:
\begin{equation}\label{DBI-Einstein.frame}
S_{DBI}^E\,=\,- T_4 \,\int_{{\cal M}_{10}}\,d^{10} x\,\,e^{\Phi/4}
\sqrt{- \det G^E}\,\big|\,\Omega\,\big|^E\,\,,
\end{equation}
where $G^E=e^{-\Phi/2}\,G$ is the Einstein frame metric and the modulus $\big|\,\Omega\,\big|^E$ is computed with $G^E$ (from now on we will suppress the index $E$ of $G^E$). It follows immediately from (\ref{Omega-Einstein}) that $\big|\,\Omega\,\big|^E$ is given by:
\beq
|\Omega|^{E} \, = \, \frac{N_f}{16 \pi^2} \, \frac{e^{9\Phi/4}}{R^2}\,
\,\frac{\delta(r-r_Q)}{r^2 \, |{\cal Z}|} \,\,.
\eeq
The total action of the system is just the one of type IIA supergravity plus the DBI+WZ action of the flavor branes. The corresponding equation of motion for the dilaton is:

\begin{equation}\label{dilaton-flavor}
\square \, \Phi \,= \, \frac{1}{4 \cdot 4!}\,e^{\Phi/2}\, F_4^2 +
2\kappa_{10}^2\,T_4\,{e^{\Phi/4}\over 4}\,|\Omega|^{E}
\, ,
\end{equation}
Again, one can show that (\ref{dilaton-flavor}) is satisfied as a consequence of the system  (\ref{flavored-BPSsystem}). The verification of this fact is straightforward (but tedious) and will not be detailed here. Let us just mention the fact that one has to  evaluate the left-hand side of (\ref{dilaton-flavor})  by computing the derivatives of the first-order BPS equations in (\ref{flavored-BPSsystem}). In this process one generates some terms in which the Heaviside function $\Theta(r-r_Q)$ is differentiated    and, therefore, the Dirac delta function $\delta(r-r_Q)$ is produced. These terms match precisely the one containing $\big|\,\Omega\,\big|^E$ in (\ref{dilaton-flavor}), while the remaining ones correspond to the $F_4^2$ term.

The Einstein equations in the Einstein frame are:
\bear
&&R_{MN}-{1\over 2}\,G_{MN}R\,=\,{1 \over 2}\left[\partial_{M}\Phi\partial_{N}\Phi-{1 \over 2}\,G_{MN}(\partial \Phi)^2 \right]\,+\,\rc\rc
&&\qquad\qquad\qquad
+\,{1\over 2\cdot 4!}e^{\Phi/2}\left[4(F_4^{\,2})_{M N}-\frac{1}{2}\,G_{MN}F_4^{\,2} \right]
\,+\,T_{MN}\,\,,
\label{Einstein}
\eear
where $T_{MN}$ is the energy-momentum tensor of the smeared flavor brane, defined as:
\beq
T_{MN}\,=\,-{2\kappa_{10}^2\over \sqrt{-G}}\,{\delta S_{DBI}^E\over \delta G^{MN}}\,\,.
\eeq
Taking into account the form of  $S_{DBI}$ in (\ref{DBI-Einstein.frame})  we arrive at the following expression for $T_{MN}$ in flat components:
\beq
T_{\bar M \bar N}\,=\,2 \kappa_{10}^2 T_4 \, {e^{\Phi/4} \over 2}\,
\Big[\,\eta_{\bar M \bar N}\,\big|\,\Omega\,\big|^{E}\,-\,{1\over 4!}\,{1\over
 \big|\,\Omega\,\big|^{E}}
\,\,(\Omega^{\,2})_{\bar M \bar N}\,\Big]\,\,.
\label{TMN}
\eeq
From this expression one readily gets the explicit values of the different components of $T_{\bar M\bar N}$, which are:
\bear
&&-T_{00}\,=\,T_{11}\,=\,T_{22}\,=\,T_{88}\,=\,T_{99}\,=\frac{R}{8 } \,
 \frac{N_f}{2 N_c}\,e^{5\Phi/2}\, \frac{\delta(r-r_Q)}{2\,r^2 \, {\cal Z}}\,,\rc\rc
&&\,T_{33}\,=\,T_{44}\,=\,T_{55}\,=\,T_{66}\,=\,T_{77}\,=\,0\,\,.
\eear
By using these values one can verify, following the same strategy used to prove (\ref{dilaton-flavor}),
 that the Einstein   equations \eqref{Einstein} are satisfied as a consequence of the first-order equations (\ref{flavored-BPSsystem}).

\section{Supersymmetric embeddings}
\label{embeddings}

In this appendix we will characterize a class of supersymmetric embeddings of D4-branes in the backgrounds  of sections \ref{unflavored} and \ref{flavoredbackground}. Our main tool will be kappa symmetry \cite{swedes}, which states that the supersymmetric embeddings of the D4-brane are those that satisfy the condition $\Gamma_{\kappa}\,\epsilon\,=\,\epsilon$, where $\epsilon$ is a Killing spinor of the background and $\Gamma_{\kappa}$ is a matrix which depends on the embedding. To write the precise form of $\Gamma_{\kappa}$, let us define the induced Dirac matrices on the D4-brane worldvolume as 
$\gamma_a=\partial_a X^M\,E_{M}^{\bar M}\,\Gamma_{\bar M}$, where 
$X^M(\xi^a)$ are the functions that parameterize  the embedding and $E_{M}^{\bar M}$ are the vielbein coefficients of ten-dimensional metric. Then, 
when the worldvolume gauge field $F$ is zero, the matrix
$\Gamma_{\kappa}$ for the D4-brane is \cite{bbs}:
\begin{equation}
\Gamma_{\kappa}\,=\,{1\over 5!}\,\,{1\over \sqrt{-\det \hat G_5}}\,\,\,
\Gamma_{11} \,\, \epsilon^{a_1\cdots a_5}\,\,\gamma_{a_1\cdots a_5} \, \ ,
\label{Gamma-kappa}
\end{equation}
where $\gamma_{a_1\cdots a_5}$ denotes the antisymmetrized product of the induced
matrices, $ \hat G_5$ is the induced metric on the D4-brane worldvolume and
$\Gamma_{11}$ is the chiral matrix in ten dimensions. Let us choose, as in (\ref{wv-coordinates}), $x^0$,  $x^1$,  $x^2$, $\sigma$ and $\psi$ as worldvolume coordinates and let us consider embeddings  as in (\ref{embedding-ansatz}), in which $\tilde\theta$ and $\tilde\phi$ depend on $(\sigma, \psi)$ and the remaining coordinates are constant. Then, the kappa symmetry matrix 
(\ref{Gamma-kappa}) takes the following form:
\begin{equation}
\Gamma_{\kappa}\,=\,{1\over \sqrt{- \det \hat G_5}}\,\,\Gamma_{11}\,
\gamma_{x^0 x^1 x^2  \sigma\psi} \, ,
\end{equation}
where the induced Gamma matrices are:
\begin{eqnarray}
&& \gamma_{x^{0,1,2}}=e^{\Phi}\Gamma_{0,1,2} \ ,  \nonumber \\
&& \gamma_{\sigma}=R\,e^{\Phi}{\cal Z}^{1/2}\left[\partial_{\sigma}\tilde{\theta} \, \Gamma_{3} 
+ \sin{\tilde{\theta}}\,\partial_{\sigma}\tilde{\phi} \, \Gamma_{4}\right]+ \frac{1}{e^{\Phi}{\cal Z}^{1/2}}\left[\Gamma_5+\sigma \, \cos{\tilde{\theta}}\,\partial_{\sigma}\tilde{\phi} \, \Gamma_{6}\right]\,\,,\\
&& \gamma_{\psi}=R\,e^{\Phi}{\cal Z}^{1/2}\left[\partial_{\psi}\tilde{\theta} \, \Gamma_{3} + 
\sin{\tilde{\theta}}\,\partial_{\psi}\tilde{\phi} \, \Gamma_{4}\right]+ \frac{\sigma}{e^{\Phi}{\cal Z}^{1/2}}\left[1+\cos{\tilde{\theta}}\,\partial_{\psi}\tilde{\phi} \right]\Gamma_{6}\,\,. \nonumber
\end{eqnarray}
To find those embeddings that are kappa symmetric and preserve the
same amount of supersymmetry as the original background, we should
compute the action of the antisymmetrized product $\gamma_{x^0\,x^1\,x^2\sigma\psi}$
on the spinor. Taking into account the projections (\ref{projections}), 
we have:
\begin{equation}
e^{-3\Phi}\Gamma_{11} \,\gamma_{x^0\,x^1\,x^2\sigma\psi}\,\epsilon=
\left[c_I + c_{39}\Gamma_{39} + c_{49}\Gamma_{49}\right]\epsilon\,\,,
\label{Gammas-on-epsilon}
\end{equation}
where the coefficients appearing on the right-hand side
of the above equation are:
\begin{eqnarray}
&& c_I=\frac{\sigma}{e^{2\Phi}{\cal Z}}\left[1+\sigma \,
\cos{\tilde{\theta}}\,\partial_{\psi}\tilde{\phi} \right] +
R^2 \,e^{2\Phi}{\cal Z}\sin{\tilde{\theta}}\left[\partial_{\psi}\tilde{\theta} \,
\partial_{\sigma}\tilde{\phi} - \partial_{\sigma}\tilde{\theta} \,
\partial_{\psi}\tilde{\phi} \right] \ ,  \nonumber \\
&& c_{39}= R \partial_{\psi}\tilde{\theta} - R\, \sigma \,
\sin{\tilde{\theta}}\,\partial_{\sigma}\tilde{\phi} \ ,  \nonumber \\
&& c_{49}= R\sin{\tilde{\theta}} \partial_{\psi}\tilde{\phi} +
R\, \sigma \left[ \partial_{\sigma}\tilde{\theta} + \cos{\tilde{\theta}}\left(\partial_{\psi}\tilde{\phi} \, \partial_{\sigma}\tilde{\theta} - \partial_{\sigma}\tilde{\phi} \, \partial_{\psi}\tilde{\theta} \right) \right] \ .
\end{eqnarray}
The right-hand side of (\ref{Gammas-on-epsilon}) should contain only the term with the identity matrix if we want to satisfy the  $\Gamma_{\kappa}\,\epsilon\,=\,\epsilon$ condition for any Killing spinor of the background. Thus, we must demand that:
\begin{equation}
c_{39}=c_{49}=0\,\,,
\end{equation}
which leads us to the following system of PDE's:
\begin{eqnarray} \label{Higgs-BPS}
&& \partial_{\psi}\tilde{\theta} - \sigma \, \sin{\tilde{\theta}}\,\partial_{\sigma}\tilde{\phi}=0 \ , \nonumber \\
&& \sin{\tilde{\theta}} \partial_{\psi}\tilde{\phi} + \sigma \left[ \partial_{\sigma}\tilde{\theta} + \cos{\tilde{\theta}}
\left(\partial_{\psi}\tilde{\phi} \, \partial_{\sigma}\tilde{\theta} - \partial_{\sigma}\tilde{\phi} \, \partial_{\psi}\tilde{\theta} \right) \right]=0 \ .
\end{eqnarray}
The general solution of the system (\ref{Higgs-BPS}) was obtained in ref. \cite{Arean:2008az}.  As mentioned in subsection \ref{Higgsbranch}, if one defines the two complex variables $\zeta_1$ and $\zeta_2$ as in (\ref{def-zetas}), any holomorphic function $\zeta_1=f(\zeta_2)$ solves (\ref{Higgs-BPS}). The corresponding embedding preserves the eight supersymmetries of the background.

\section{Additional supergravity backgrounds}
\label{additonal-backgrounds}

In this appendix we find additional supergravity solutions for our unflavored setup. First of all we will show that it is possible to find solutions for the unflavored BPS system of equations \eqref{no-flavor-BPSsystem} that are  simpler  than those coming from gauged supergravity. Later on, in subsection \ref{nonrelativistic}, after performing a series of duality transformations to the unflavored solution of section \ref{unflavored}, we will generate a background dual to a non-relativistic system in 1+1 dimensions.

Let us try to solve the unflavored BPS system  \eqref{no-flavor-BPSsystem} by means of the method of separation of variables. Accordingly, let us adopt the following ansatz for the three functions $g$, $\Phi$ and ${\cal Z}$ appearing in the background:
\beq
g\,=\,g(\sigma)\,\,,\qquad
\Phi\,=\,\Phi(r)\,\,,\qquad
{\cal Z}\,=\,{\cal Z}_1(r)\,{\cal Z}_2(\sigma)\,\,.
\label{separated-ansatz}
\eeq
It is straightforward to show that the general solution of the system (\ref{no-flavor-BPSsystem}) that has the form of the ansatz (\ref{separated-ansatz}) is:
\bear
&&g(\sigma)\,=\,c_1\, R^2\,\sqrt{c_4\,+\,c_3\,\,\sigma^2}\,\,,\rc\rc
&&e^{-4\Phi(r)}\,=\,c_3\, R^2\,\,\Big[\,c_2\,+\,{c_1\over r}\,\Big]^2\,\,,\rc\rc
&&{\cal Z}(r,\sigma)\,=\,\Big(c_2+{c_1\over r}\,\Big)\,\sqrt{c_4\,+\,c_3\,\,\sigma^2}\,\,,
\label{gPhiZ-new}
\eear
where the $c_i$'s are separation constants.  The ten-dimensional metric corresponding to this solution  can be obtained by plugging the values of $e^{-4\Phi}$ and ${\cal Z}$ given in \eqref{gPhiZ-new} into a  general ansatz. Proceeding in  this way we arrive at the split 6+4 metric:
\beq
ds^2_{10}\,=\,ds^2_{6}\,+\,ds^2_{4}\,\,,
\eeq
where the six-dimensional metric is independent of $\sigma$:
\beq
ds^2_6\,=\,e^{2\Phi}\,dx^2_{1,2}\,+\,e^{-2\Phi}\,
\left[\,dr^2\,+\,r^2\,d\Omega_2^2\,\right]\,\,,
\eeq
while the four-dimensional metric is independent of $r$:
\beq
ds^2_4\,=\,{R\,\sqrt{c_4\,+\,c_3\,\,\sigma^2}\over \sqrt{c_3}}\,\,
\,\left(d\tilde{\theta}^2+\sin^2\tilde{\theta}d\tilde{\phi}^2\right)\,+\,
{R\,\sqrt{c_3}\over \sqrt{c_4\,+\,c_3\,\,\sigma^2}}\,\,
\Big[\,d\sigma^2\,+\,\sigma^2\,\big(\,d\psi+\cos\tilde{\theta} d\tilde{\phi}\,\big)^2\,\Big]\,\,.\qquad\qquad
\label{ds4}
\eeq
In order to rewrite the metric \eqref{ds4} in a more familiar form, let us perform a change of variables and define a new radial variable $\zeta$, related to $\sigma$ in the following way:
\beq
\zeta^2\,=\,{4\, R\over \sqrt{c_3}} \,\,\sqrt{c_4\,+\,c_3\,\,\sigma^2}\,\,.
\label{zeta}
\eeq
Actually, if we define a new constant $a$ as:
\beq
a^4\,=\,{16 R^2 c_4\over c_3}\,\,,
\eeq
the relation that gives $\sigma$ in terms of $\zeta$ is:
\beq
\sigma\,=\,{1\over 4\,R}\,\sqrt{\zeta^4-a^4}\,\,.
\eeq
Clearly, $\zeta\ge a$, which corresponds to the range $\sigma\ge 0$. After this change of variable, one can easily prove that the metric \eqref{ds4} becomes:
\beq
ds^2_4\,={d\zeta^2\over 1\,-\,\big({a\over \zeta}\big)^4}\,+\,
{\zeta^2\over 4}\,\,\Big[\,d\tilde{\Omega}_2^2\,+\,
\Big(\,1\,-\,\Big({a\over \zeta}\Big)^4\,\Big)\,\big(\,d\psi+\cos\tilde{\theta} d\tilde{\phi}\,\big)^2\,\Big]\,\,,
\label{EH}
\eeq
which is the metric of an Eguchi-Hanson space $EH_4$ with resolution parameter $a$.  When $a=0$ the metric \eqref{EH} becomes the one corresponding to the $\mathbb{C}^2/\mathbb{Z}_2$ orbifold. Moreover, if we define the new constants $\eta$ and $Q$ as:
\beq
\eta\,\equiv\,R\,\sqrt{c_3}\, c_2\,\,,\qquad\qquad
Q\,\equiv\,R\,\sqrt{c_3}\, c_1\,\,,
\eeq
then, the warp factor $e^{-4\Phi}$ becomes:
\beq
e^{-4\Phi}\,=\,\left(\,\eta\,+\,{Q\over r}\,\right)^2\,\,.
\label{Phi-new}
\eeq
Using this result the six-dimensional part of the metric takes the form:
\beq
ds^2_6\,=\,{dx^2_{1,2}\over \eta\,+\,{Q\over r}} \,+\,
\left(\,\eta\,+\,{Q \over r}\,\right)\,
\Big[\,dr^2\,+\,r^2\,d\Omega_2^2\,\Big]\,\,.
\label{ds6}
\eeq
Notice that, for $\eta\not= 0$, the metric (\ref{ds6}) is asymptotically flat.

%%%%%%%%%%%%%%%%%%%%%%%%%%%%%%%%%%%%%%%%%%%%%%%%%%

\subsection{Non-relativistic backgrounds}
\label{nonrelativistic}

Let us now follow the procedure of ref. \cite{NR} to obtain backgrounds dual to non-relativistic systems by performing a combination of two T-dualities and a shift to a supergravity dual of a relativistic theory.  First of all, we introduce light-cone variables in the standard way, $x^{\pm} = x^0 \pm x^1$, and rewrite the initial metric and the dilaton of the unflavored type IIA background as:
\begin{eqnarray}
&& ds_{IIA}^2 = H^{-1/2}\left[-\,dx^{+}dx^{-}+(dx^2)^2+{\cal Z} R^2
\left(d\tilde{\theta}^2+\sin^2\tilde{\theta}d\tilde{\phi}^2\right)\right]\,+\,\nonumber\\
&& \quad \quad \quad \quad + \, H^{1/2}\left[dr^2+r^2d\Omega^2_2+\frac{1}{{\cal Z}}
\left(d\sigma^2+\sigma^2\left(d\psi+\cos\tilde{\theta}d\tilde{\phi}\right)^2\right)\right]\,\,,\nonumber \\
&& e^{2\Phi}=H^{-1/2}   \ .
\end{eqnarray}
Notice that we have introduced the warp factor $H$. Similarly, 
the RR potentials in these variables take the form:
\begin{eqnarray}
C_3 & = &  - \, g\,\omega_2 \wedge (d\psi+\cos\tilde{\theta} \,d\tilde{\phi}) \,\,, 
\nonumber \\
C_5 & = & -\,\frac{1}{2}\,dx^{+}\wedge dx^{-}\wedge dx^2 \wedge \left[\,{R^2 {\cal Z}
\over H}\,\tilde{\omega}_2 \,-\,{\sigma \over {\cal
Z}}\,\,d\sigma\wedge (d\psi+\cos\tilde{\theta}
\,d\tilde{\phi})\,\right] \ .
\end{eqnarray}
In the first step we perform a T-duality along the fiber direction $\psi$. In general, under T-duality
the fields in the NSNS sector form a closed set and transform amongst themselves.
Hence for these fields we may use the standard rules. In contrast, the transformation rules
for the RR sector fields involve those in the NSNS sector. The metric
and dilaton after this T-duality become:
\begin{eqnarray}
&& ds_{IIB}^2 = H^{-1/2}\left[-\,dx^{+}dx^{-}+(dx^2)^2+{\cal Z} R^2
\left(d\tilde{\theta}^2+\sin^2\tilde{\theta}d\tilde{\phi}^2\right)\right]\,+\,\nonumber\\
&& \quad \quad \quad \quad + \, H^{1/2}\left[dr^2+r^2d\Omega^2_2+\frac{1}{{\cal Z}}\,
d\sigma^2\right] + H^{-1/2}\frac{{\cal Z}}{\sigma^2}\,d\psi^2 \,\,,\nonumber \\
&& e^{2\Phi}=\frac{{\cal Z}}{\sigma^2}\,H^{-1}   \ ,
\end{eqnarray}
while the non-vanishing NSNS and RR potentials are:
\begin{eqnarray}
B_2 & = & \cos\tilde{\theta}\,d\psi \wedge d\tilde{\phi}\,\,, \nonumber \\
C_2 & = & -g \omega_2 \ , \quad C_4 = \frac{\sigma}{2{\cal Z}}\,
dx^{+} \wedge dx^{-} \wedge dx^2 \wedge d\sigma\,\,, \nonumber \\
C_6 & = & -\,\frac{R^2 {\cal Z}}{2H}\,
dx^{+} \wedge dx^{-} \wedge dx^2 \wedge \tilde{\omega}_2\wedge d\psi \ .
\end{eqnarray}
In the second step we perform a coordinate shift along the
light-cone coordinate $x^{-}$ of the form:
\begin{equation}
x^{-} \rightarrow x^{-} + \gamma \psi \quad \longrightarrow \quad
dx^{-} \rightarrow dx^{-}+\gamma d\psi \ .
\end{equation}
The corresponding metric and the dilaton become:
\begin{eqnarray}
&& ds_{IIB}^2 = H^{-1/2}\left[-\,dx^{+}dx^{-}-\gamma dx^{+}d\psi+(dx^2)^2+{\cal Z} R^2
\left(d\tilde{\theta}^2+\sin^2\tilde{\theta}d\tilde{\phi}^2\right)\right]\,+\nonumber\\
&& \quad \quad \quad \quad + \, H^{1/2}\left[dr^2+r^2d\Omega^2_2+\frac{1}{{\cal Z}}\,
d\sigma^2\right] + H^{-1/2}\frac{{\cal Z}}{\sigma^2}\,d\psi^2\,\,, \nonumber \\
&& e^{2\Phi}=\frac{{\cal Z}}{\sigma^2}\,H^{-1}   \ .
\end{eqnarray}
while the non-vanishing NSNS and RR potentials are:
\begin{eqnarray}
B_2 & = & \cos\tilde{\theta}\,d\psi \wedge d\tilde{\phi}\,\,, \nonumber \\
C_2 & = & -g \omega_2 \ , \quad C_4 = \frac{\sigma}{2{\cal Z}}\,
dx^{+} \wedge \left(dx^{-}+\gamma d\psi \right) \wedge dx^2 \wedge d\sigma\,\,, \nonumber \\
C_6 & = & -\,\frac{R^2 {\cal Z}}{2H}\,
dx^{+} \wedge dx^{-} \wedge dx^2 \wedge \tilde{\omega}_2\wedge d\psi \ .
\end{eqnarray}
Finally we perform another T-duality along $\psi$ and return to a type
IIA background. Now the metric and the dilaton become:
\begin{eqnarray}
&& ds_{IIA}^2 = H^{-1/2}\left[-\,dx^{+}\left(dx^{-}+\frac{\gamma^2}{4}
\frac{\sigma^2}{{\cal Z}}\,dx^{+}\right)+(dx^2)^2+{\cal Z} R^2
\left(d\tilde{\theta}^2+\sin^2\tilde{\theta}d\tilde{\phi}^2\right)\right]\,\,+\nonumber\\
&& \quad \quad \quad \quad + \, H^{1/2}\left[dr^2+r^2d\Omega^2_2+\frac{1}{{\cal Z}}
\left(d\sigma^2+\sigma^2\left(d\psi+\cos\tilde{\theta}d\tilde{\phi}\right)^2\right)\right] \,\,,\nonumber \\
&& e^{2\Phi}= H^{-1/2}   \ .
\end{eqnarray}
while the non-vanishing NSNS and RR potentials are:
\begin{eqnarray}
B_2 & = &  \frac{\gamma}{2}\frac{\sigma^2}{{\cal Z}}\,dx^{+}
\wedge (d\psi + \cos\tilde{\theta} d\tilde{\phi}) \,\,,\nonumber \\
C_3 & = &  - \, g\,\omega_2 \wedge (d\psi+\cos\tilde{\theta} \,d\tilde{\phi})
+ \frac{\gamma}{2}\frac{\sigma}{{\cal Z}}\,dx^{+}
\wedge dx^2 \wedge d\sigma\,\,, \nonumber \\
C_5 & = & -\,\frac{1}{2}\,dx^{+}\wedge dx^{-}\wedge dx^2 \wedge \left[\,{R^2 {\cal Z}
\over 2H}\,\tilde{\omega}_2 \,-\,{\sigma \over {\cal
Z}}\,\,d\sigma\wedge (d\psi+\cos\tilde{\theta}
\,d\tilde{\phi})\,\right] \ .
\end{eqnarray}
The corresponding field strengths of these potentials are $H_3=dB_2$,
$F_4=dC_3$. By computing the exterior derivatives
and using the BPS equations we have:
\begin{eqnarray}
&& \nonumber H_3 \,=\,{\gamma\over 2}\,\Bigg[\,{g \sigma^2
\over {\cal Z}^2 r^2 R^2}\,dr \,+ \,
{2\sigma\over {\cal Z}}\,\Big(\,1\,-\,
{\sigma^2 H \over 2 {\cal Z}^2 R^2} \,d\sigma\,\Big)\,\Bigg]
\wedge dx^+\wedge \left(\,d\psi+\cos\tilde{\theta} d\tilde{\phi}\,\right)\,+
\,{\gamma\over 2}\,
{\sigma^2\over {\cal Z}}\,dx^+\wedge \tilde{\omega}_2 \ , \\
&&F_4\,= - \, dg \wedge \omega_2 \wedge \left(\,d\psi+\cos\tilde{\theta} d\tilde{\phi}\,\right)
+ \, g \,  \omega_2 \wedge \tilde{\omega}_2
+\,{\gamma\over 2}\,{\sigma g\over r^2 {\cal Z}^2 R^2}\,
dx^+\wedge dx^2 \wedge dr \wedge d\sigma \ .
\end{eqnarray}
As a consistency check we can verify that these forms satisfy their equations of motion:
\begin{equation}
d \Big( e^{-2\Phi}\, {}^* H_3 \Big )\,=  \frac{1}{2}\,F_4 \wedge F_4 \ , \qquad\qquad
d \Big (\, {}^* F_4 \Big )\,=\, H_3 \wedge F_4 \ .
\end{equation}

\section{Entanglement entropy}
\label{entanglement}

In quantum field theory the entanglement entropy between two complementary spatial regions $A$ and $B$ is defined as the entropy seen by an observer in $A$ who does not have access to the degrees of freedom of $B$. The holographic computation consists of finding the eight-dimensional surface $\Sigma$ with minimal area such that its boundary coincides with the boundary of $A$.  Then, the entanglement entropy between $A$ and
its complementary region $B$ is given by the integral \cite{Ryu}:
\begin{equation}
S=\frac{1}{4G_{10}}\int_{\Sigma} d^8 \xi \: e^{-2\Phi}\sqrt{\hat{G}_{8}}
\ ,
\end{equation}
where $G_{10}$ is the ten-dimensional Newton constant, given by $G_{10}=8\pi^6\alpha'^4 g_s^2$ and $\hat G_8$ is the induced metric on $\Sigma$ (see also \cite{Klebanov:2007ws}). We will consider a constant time surface $\Sigma$, obtained by minimizing $S$ over all surfaces that approach the boundary of $A$ at the boundary of the ten-dimensional  bulk manifold and that are extended along the remaining spatial directions. The surface we consider is not going
to touch the IR region of the space, so it is feasible to use the UV
expression \eqref{UV}.
We parametrize the eight-dimensional surface in the following way:
\begin{equation}
\xi^a=(x, x_2, \tilde{\theta}, \tilde{\phi},
\hat{\alpha}, \theta, \phi, \psi) \ ,
\end{equation}
with $u=u(x)$. By computing the induced metric we end up
with the following expression for $S$:
\begin{equation}
S=\frac{2\,\pi^3R^5}{3\,G_{10}}\, \int^{\frac{l}{2}}_
{-\frac{l}{2}} dx\, u \,\sqrt{u'^2+\frac{8}{\sqrt{{\cal Z}_{*}}}\left(\frac{u}{R}\right)^3} \ .
\end{equation}
Since the function $S$ does not depend explicitly on $x$,
the Euler-Lagrange equation derived from $S$ can be integrated with
the result being the following:
\begin{equation}
\frac{u^4}{ \sqrt{u'^2+\frac{8}{\sqrt{{\cal Z}_{*}}}\left(\frac{u}{R}\right)^3}} = u_0^{5/2}\,\,
\frac{R^{3/2}\,{\cal Z}_{*}^{1/4}}{2\sqrt{2}}\ ,
\end{equation}
where $u_0$ is the minimal value of $u$. From this expression we can
obtain $u'$ as a function of $u$:
\begin{equation} \label{u'}
u'=\pm \sqrt{\frac{8}{R^3 {\cal Z}_{*}^{1/2}}}\,\, u^{3/2}\,\,
\sqrt{\frac{u^5}{u_0^5}-1} \ .
\end{equation}
Now we can compute the length $l$ as a function of the turning point
$u_0$ of the holographic coordinate:
\begin{equation} \label{length}
l=2 \int_{u_0}^{\infty}\,{du\over |\,u'\,|}\,=\,
\sqrt{2\pi \sqrt{{\cal Z}_{*}}\,R^3} \,\,\,
\frac{\Gamma(\frac{3}{5})}{\Gamma(\frac{1}{10})} \, \frac{1}{\sqrt{u_0}} \ .
\end{equation}
We can use \eqref{u'} to eliminate $u'$ in the entropy functional $S$.
The resulting integral is divergent if the upper limit is infinity.
For this reason we regulate it by integrating up to some value $u_{\infty}$ of $u$ and
we have:
\begin{equation} \label{S}
S=\frac{2\,\pi^3\,R^5}{3\,G_{10}}\,\, u_0^2\, \int_1^{\frac{u_{\infty}}{u_0}}
\frac{\xi^{7/2}}{\sqrt{\xi^5-1}} \: d\xi  \ .
\end{equation}
The integral appearing on the right hand side of the
above equation takes the value:
\begin{equation}
\int_1^{\frac{u_{\infty}}{u_0}}
\frac{\xi^{7/2}}{\sqrt{\xi^5-1}} \: d\xi = -\frac{\sqrt{\pi}}{2}
\frac{\Gamma(\frac{3}{5})}{\Gamma(\frac{1}{10})}-\frac{1}{5}\:B\left[\frac{u_0^5}
{u_{\infty}^5},-\frac{2}{5},\frac{1}{2} \right] \ ,
\end{equation}
where $B$ is the symbol for the incomplete Beta function. Taking the limit
of the above expression when $u_{\infty}\rightarrow \infty$ we 
obtain finite and divergent terms:
\begin{equation}
\int_1^{\frac{u_{\infty}}{u_0}}
\frac{\xi^{7/2}}{\sqrt{\xi^5-1}} \: d\xi \approx -\frac{\sqrt{\pi}}{2}
\frac{\Gamma(\frac{3}{5})}{\Gamma(\frac{1}{10})}+
\frac{1}{2}\left(\frac{u_{\infty}}{u_0}\right)^2 + {\cal O}\left(\frac{u_0}{u_{\infty}}\right)^2\ .
\end{equation}
Plugging the finite part of this result in \eqref{S} and using
\eqref{length} to express $u_0$ in terms of $l$ we have:
\begin{equation} \label{entropy}
S^{finite}(l)=-\left(2\pi R_{10}\right) V_{\infty} \frac{2^5 \sqrt{\pi}}{3} \left[
\frac{\Gamma(\frac{3}{5})}{\Gamma(\frac{1}{10})}\right]^5 \frac{N_c^3}{l^4}  \ ,
\end{equation}
where $R_{10}=g_s \alpha'^{1/2}$ and $V_{\infty}$ is the volume of a sphere
of radius $R {\cal Z}_{*}^{1/2}$ along
which the D4-branes are wrapped:
\begin{equation}
V_{\infty}= 4\,\pi R^2 \, {\cal Z}_{*} \ .
\end{equation}
The expression (\ref{entropy}) just obtained coincides with the one found in \cite{Klebanov:2007ws} for an M5-brane compactified  in a two sphere of volume $V_{\infty}$. This result was, of course, to be expected, given the UV behavior of our metric.

%%%%%%%%%%%%%%%%%%%%%%%%%%%%%%%%%%%%%%%%%%%%%%%%%%%%%%%%%%%%%%%%%%%%%

%%%%%%%%%%%%%%%%%%%%%%%%%%%%%%%%%%%%%%%%%%%%%%%%%%%%%%%%%%%%%%%%%%%%%%%%%%

\end{document}